\documentclass[longauth]{aa}

\usepackage{color}
\definecolor{black}{rgb}{0,0,0}
\definecolor{red}{rgb}{1,0,0}			
\definecolor{darkblue}{rgb}{0,0,0.7}

\usepackage{longtable,array}
\usepackage{graphicx}
\usepackage{txfonts}

\begin{document}

   \title{Pluto's lower atmosphere and pressure evolution from ground-based stellar occultations, 1988-2016}

      \subtitle{}

   \author{E. Meza\inst{1}
          \fnmsep\thanks{%
          Partly based on observations made with the 
          Ultracam camera at the Very Large Telescope (VLT Paranal), under program ID 079.C-0345(F), 
          the ESO camera NACO at VLT, under program IDs 079.C-0345(B), 089.C-0314(C) and 291.C- 5016, 
          the ESO camera ISAAC at VLT under program ID 085.C-0225(A),  
          the ESO camera SOFI at NTT Paranal,  under program ID 085.C-0225(B),
          the WFI camera at 2.2m La Silla, under program ID's
          079.A-9202(A), 075.C-0154, 077.C-0283, 079.C-0345,  088.C-0434(A),
          089.C-0356(A), 090.C-0118(A) and 091.C-0454(A),
          the Laborat\'orio Nacional de Astrof\'{\i}sica (LNA), Itajub\'a - MG, Brazil,
          the Southern Astrophysical Research (SOAR) telescope, and the
          the Italian Telescopio Nazionale Galileo (TNG).
          }%
          \and
          B. Sicardy\inst{1}
          \and
          M. Assafin\inst{2,3}
          \and
          J.~L. Ortiz\inst{4}
          \and
         T. Bertrand\inst{5}
         \and
         E. Lellouch\inst{1},
         J. Desmars\inst{1}
         \and
         F. Forget\inst{6}
         \and
         D. B\'erard\inst{1}, 
         A. Doressoundiram\inst{1}, 
         J. Lecacheux\inst{1}
% ...
          }

   \institute{LESIA, Observatoire de Paris, PSL Research University, CNRS, Sorbonne Universit\'e, 
             Univ. Paris Diderot, Sorbonne Paris Cit\'e \\
             \email{Bruno.Sicardy@obspm.fr}
             \and 
             Observat\'orio do Valongo/UFRJ, Ladeira Pedro Antonio 43, Rio de Janeiro, RJ 20080-090, Brazil
              \and 
             Observat\'orio Nacional/MCTIC, Laborat\'orio Interinstitucional de e-Astronomia-LIneA and INCT do e-Universo,
             Rua General Jos\'e Cristino 77, Rio de Janeiro CEP 20921-400, Brazil            
             \and 
             Instituto de Astrof\'{\i}sica de Andaluc\'{\i}a (IAA-CSIC). Glorieta de la Astronom\'{\i}a s/n. 18008-Granada, Spain     
             \and 
             National Aeronautics and Space Administration (NASA), Ames Research Center, Space Science Division, 
             Moffett Field, CA 94035, USA
             \and 
             Laboratoire de M\'et\'eorologie Dynamique, IPSL, Sorbonne Universit\'e, UPMC Univ. Paris 06, CNRS, 
             4 place Jussieu, 75005 Paris, France
% ...
}
 
\date{Received Sept:19, 2018; accepted Mar:01, 2019}


  \abstract
   {%
   Pluto's tenuous nitrogen (N$_2$) atmosphere undergoes strong seasonal effects due to 
   high obliquity and orbital eccentricity, and has been recently (July 2015) observed by the New Horizons spacecraft.
   }%
   {%
   Goals are 
   $(i)$ construct a well calibrated record of the seasonal evolution of surface pressure on Pluto and 
   $(ii)$ constrain the structure of the lower atmosphere using a central flash observed in 2015.
  }%
   {%
   Eleven stellar occultations by Pluto observed between 2002 and 2016 are used to retrieve 
   atmospheric profiles (density, pressure, temperature) between $\sim$5~km and $\sim$380~km altitude levels 
   (i.e. pressures from $\sim$10~$\mu$bar to 10~nbar).
   }%
   {%
   $(i)$ Pressure has suffered a monotonic increase from 1988 to 2016, that is compared to a seasonal volatile transport model,
   from which tight constraints on a combination of albedo and emissivity of N$_2$ ice are derived.
   $(ii)$ A central flash observed on 2015 June 29 is consistent with New Horizons REX profiles, provided that 
   (a) large diurnal temperature variations (not expected by current models) occur over Sputnik Planitia and/or 
   (b) hazes with tangential optical depth $\sim$0.3 are present 
   at 4-7 km altitude levels and/or 
   (c) the nominal REX density values are overestimated by an implausibly large factor of $\sim$20\% and/or 
   (d) higher terrains block part of the flash in the Charon facing hemisphere. 
   }%
   {}


  \maketitle
%

\section{Introduction}
 
Pluto's tenuous atmosphere was glimpsed during a 
ground-based stellar occultation observed on 1985 August 19 \citep{bro95},
and fully confirmed on 1988 June 09 during another occultation \citep{hub88,ell89,mil93}
that provided the main features of its structure:
temperature, composition, pressure, density, see the review by \cite{yel97}.

Since then, Earth-based stellar occultations 
have been quite an efficient method to study Pluto's atmosphere.
It yields, in the best cases, information from 
a few kilometers above the surface (pressure $\sim$10~$\mu$bar) 
up to  380~km altitude ($\sim$10~nbar).
As Pluto moved in front of the Galactic center, 
the yearly rate of stellar occultations dramatically increased during the 2002-2016 period,
yielding a few events per year that greatly improved our knowledge of Pluto's atmospheric structure and evolution.

Ground-based occultations also provided a decadal monitoring of the atmosphere.
Pluto has a large obliquity 
($\sim 120\degr$, the axial inclination to its orbital plane) 
and high orbital eccentricity (0.25) that
takes the dwarf planet from 29.7 to 49.3~AU during half of its 248-year orbital period.
Northern spring equinox occurred in January 1988 and perihelion occurred soon after, in September 1989.
Consequently, our survey monitored Pluto as it receded from the Sun 
while exposing more and more of its northern hemisphere to sunlight. 
More precisely, as of 2016 July 19 (the date of the most recent occultation reported here), 
Pluto's heliocentric distance has increased by a factor of 1.12 since perihelion, 
corresponding to a decrease of about 25\% in average insolation.
Meanwhile, the subsolar latitude has gone from zero degree at equinox to $54\degr$ north in July 2016.
In this context, dramatic seasonal effects are expected, and observed.

Another important aspect of ground-based occultations is that 
they set the scene for the NASA New Horizons mission (NH hereafter)
that flew by the dwarf planet in July 2015 \citep{ste15}. 
A fruitful and complementary comparison between the ground-based and NH results ensued --
another facet of this work.

Here we report results derived from eleven Pluto stellar occultations observed between 2002 and 2016,
five of them yet unpublished, as mentioned below.
We analyze them in a unique and consistent way.
Including the 1988 June 09 occultation results, and using 
the recent surface ice inventory provided by NH, 
we constrain current seasonal models of the dwarf planet.
Moreover, a central flash observed during the 2015 June 29 occultation is used to compare 
Pluto's lower atmosphere structure derived from the flash 
with profiles obtained by the Radio Science EXperiment instrument on board of NH (REX hereafter)
below an altitude of about 115~km

Observations, data analysis and primary results are presented in Section~\ref{sec_data_analysis}.
Implications for volatile transport models are discussed in Section~\ref{sec_pres_evolution}.
The analysis of the 2015 June 29 central flash is detailed in Section~\ref{sec_lower_atmo}, 
together with its consequences for Pluto's lower atmosphere structure.
Concluding remarks are provided in Section~\ref{sec_conclusion}.

\section{Observations and data analysis}
\label{sec_data_analysis}

\subsection{Occultation campaigns}

Table~\ref{tab_sites} lists the circumstances of all the 
Pluto stellar occultation campaigns that our group have organized between 2002 and 2016.
The first part of this table lists the eleven events that were used in the present work.
In a second part of the table, we list other campaigns that were not used, because
the occultation light curves had insufficient signal-to-noise-ratio and/or
because of deficiencies in the configuration of the occulting chords
(grazing chords or single chord)
and as such, do not provide relevant measurements of the atmospheric pressure.

Details on the prediction procedures can be found in \citealt{ass10,ass12,ben14}.
Some of those campaigns are already documented and analyzed in previous publications, 
namely the 
2002 July 20, 
2002 August 21, 
2007 June 14, 
2008 June 22, 
2012 July 18, 
2013 May 04 and
2015 June 29
events. 
They were used to constrain 
Pluto's global atmospheric structure and evolution \citep{sic03,dia15,fre15,olk15,sic16},
the structure and composition (CH$_4$, CO and HCN abundances) of the lower atmosphere 
by combination with spectroscopic IR and sub-mm data \citep{lel09,lel15,lel17},
the presence of gravity waves \citep{toi10,fre15} and
Charon's orbit \citep{sic11}.
Finally, one campaign that we organized is absent from Table~\ref{tab_sites} (2006 April 10).
It did not provide any chord on Pluto, 
but was used to put an upper limit of Pluto's rings \citep{boi14}.

Note that we include here five more (yet unpublished) data sets obtained on the following dates:
2008 June 24,
2010 February 14,
2010 June 04,
2011 June 04 and
2016 July 19.

\begin{table}
\caption{Adopted physical parameter}
\label{tab_param}
\centering
\begin{tabular}{ll}
\hline
Pluto's mass\tablefootmark{1}                               & $GM_P=  8.696 \times 10^{11}$ m$^3$ sec$^{-2}$ \\
Pluto's radius\tablefootmark{1}                               & $R_P= 1187$~km \\
N$_2$ molecular mass                                      & $\mu= 4.652 \times 10^{-26}$ kg \\
N$_2$ molecular                                               & $K = 1.091 \times 10^{-23}$ \\ 
refractivity\tablefootmark{2}                                 &  $+ (6.282 \times 10^{-26}/\lambda_{\rm \mu m}^2)$ cm$^3$ molecule$^{-1}$   \\
Boltzmann constant                                              & $k= 1.380626 \times 10^{-23}$ J K$^{-1}$ \\
Pluto pole position\tablefootmark{3}                  & $\alpha_{\rm p}$= 08h  52m 12.94s \\
(J2000)                                                               & $\delta_{\rm p}$= -06d 10' 04.8" \\
\hline
\end{tabular}
\tablefoot{%
\tablefoottext{1}{\cite{ste15}, where $G$ is the constant of gravitation.}
\tablefoottext{2}{\cite{was30}.}
\tablefoottext{3}{\cite{tho08}.}
}
\end{table}

\begin{figure}[!t]
\centering
\includegraphics[totalheight=6cm,angle=0, trim=0 0 0 0]{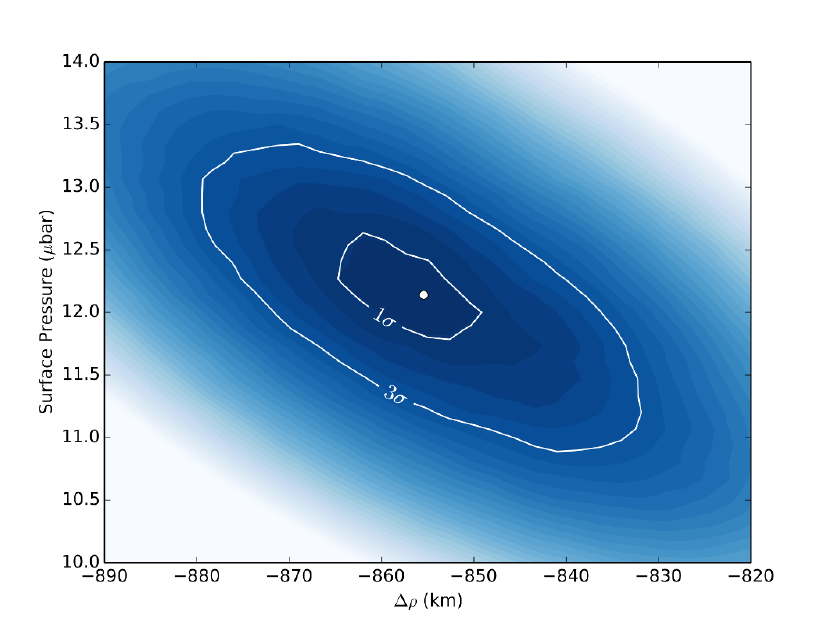}
\caption[$\chi^{2}$-map for the simultaneous fit to the 2016 July 19 Pluto light curves]{
An example of $\chi^{2}(\Delta \rho,p_{\rm surf})$ map 
derived from the simultaneous fit to the light curves obtained 
during the 2016 July 19 occultation. 
The quantity $\Delta \rho$ is Pluto's ephemeris offset (expressed in kilometers) 
perpendicular to the apparent motion of the dwarf planet, as projected in the sky plane. 
The other parameter ($p_{\rm surf}$) is the surface pressure of the DO15 atmospheric model.
The white dot marks the best fit, where the minimum value $\chi^2_{\rm min}$ of $\chi^2$ is reached.
The value $\chi^{2}_{\rm min}= 4716$, using 4432 data points, 
indicates a satisfactory fit with a $\chi^{2}$ per degree of freedom 
of $\chi^{2}_{\rm dof} \sim$~4716/4432 $\sim$~1.06.
The best fit corresponds to $p_{\rm surf} =12.04 \pm 0.41$ $\mu$bar (1-$\sigma$ level). 
The error bar is derived from the 1-$\sigma$ curve that delineates the $\chi^2_{\rm min} + 1$ level.
The 3-$\sigma$ level curve (corresponding  to the $\chi^2_{\rm min} + 9$ level) is also shown.
}
\label{fig_chi2_19jul16}
\end{figure}

\subsection{Light curve fitting}

For all the eleven data sets used here, 
we used the same procedure as in \cite{dia15} (DO15 hereafter) and in \cite{sic16}.
It consists of simultaneously fitting the refractive occultation light curves 
by synthetic profiles generated by a ray tracing code that uses the Snell-Descartes law.
The physical parameters adopted in this code are listed in Table~\ref{tab_param}.

Note in particular that our adopted Pluto's radius is taken from \cite{ste15},
who use a global fit to full-disk images provided by 
the Long-Range Reconnaissance Imager (LORRI) of NH to obtain $R_P= 1187 \pm 4$~km.
\cite{nim17} improve that value to $R_P=  1188.3 \pm 1.6$~km.
However, we kept the 1187~km value because it is very close to the deepest level reached
by the REX experiment, near the depression Sputnik Planitia, see Section~\ref{sec_lower_atmo}.
Consequently, it is physically more relevant here when discussing Pluto's lower atmospheric structure.
  
We assume a pure N$_2$ atmosphere, 
which is justified by the fact that the next most important species (CH$_4$) 
has an abundance of about 0.5\% \citep{lel09,lel15,gla16}, 
resulting in negligible effects on refractive occultations. 

We also assume a transparent atmosphere, which is supported by the NH findings.
As discussed in Section~\ref{sec_lower_atmo}, the tangential (line-of-sight) optical depth of hazes found by NH
for the rays that graze the surface is $\tau_T \sim 0.24$, 
with a scale height of $\sim$~50 km \citep{gla16,che17}.
As our fits are mainly sensitive to levels around 110~km (see below), this means that haze absorption
may be neglected in our ray tracing approach.
We return to this topic in Section 4.3, which considers the effect of haze absorption on the central flash,
possibly caused by the deepest layers accessible using occultations.

Moreover, we take a global spherically symmetric atmosphere, 
which is again supported by the NH results, at least above the altitude $\sim$35~km, 
see \cite{hin17} and Fig.~\ref{fig_n_r_rex_compa_bs}.
This is in line with Global Climate Models (GCMs), 
which predict that wind velocities in the lower atmosphere should not exceed 
$v \sim$1-10~m~s$^{-1}$ \citep{for17}.
If uniform, this wind would create an equator to pole radius difference of the corresponding
isobar level of at most $\Delta r \sim (R_P v)^2/4GM_P < 0.1$~km, 
using Eq.~7 of \cite{sic06} and the values in Table~\ref{tab_param}.
This expected distortion is too small to significantly affect our synthetic profiles.

Finally, the temperature profile $T(r)$ is taken constant. 
Here, the radius $r$ is counted from Pluto's center, while
Pluto's radius found by NH is 1187~km (Table~\ref{tab_param}).
This will be the reference radius from which we calculate altitudes.
Fixing the pressure at a prescribed level (e.g. the surface)
then entirely defines the density profile $n(r)$ 
to within a uniform scaling factor for all radii $r$, 
using the ideal gas equation, hydrostatic equilibrium assumption,
and accounting for the variation of gravity with altitude.

\begin{figure*}[!t]
\centerline{
\includegraphics[width=\paperwidth,trim=0 30 0 0,angle=0]{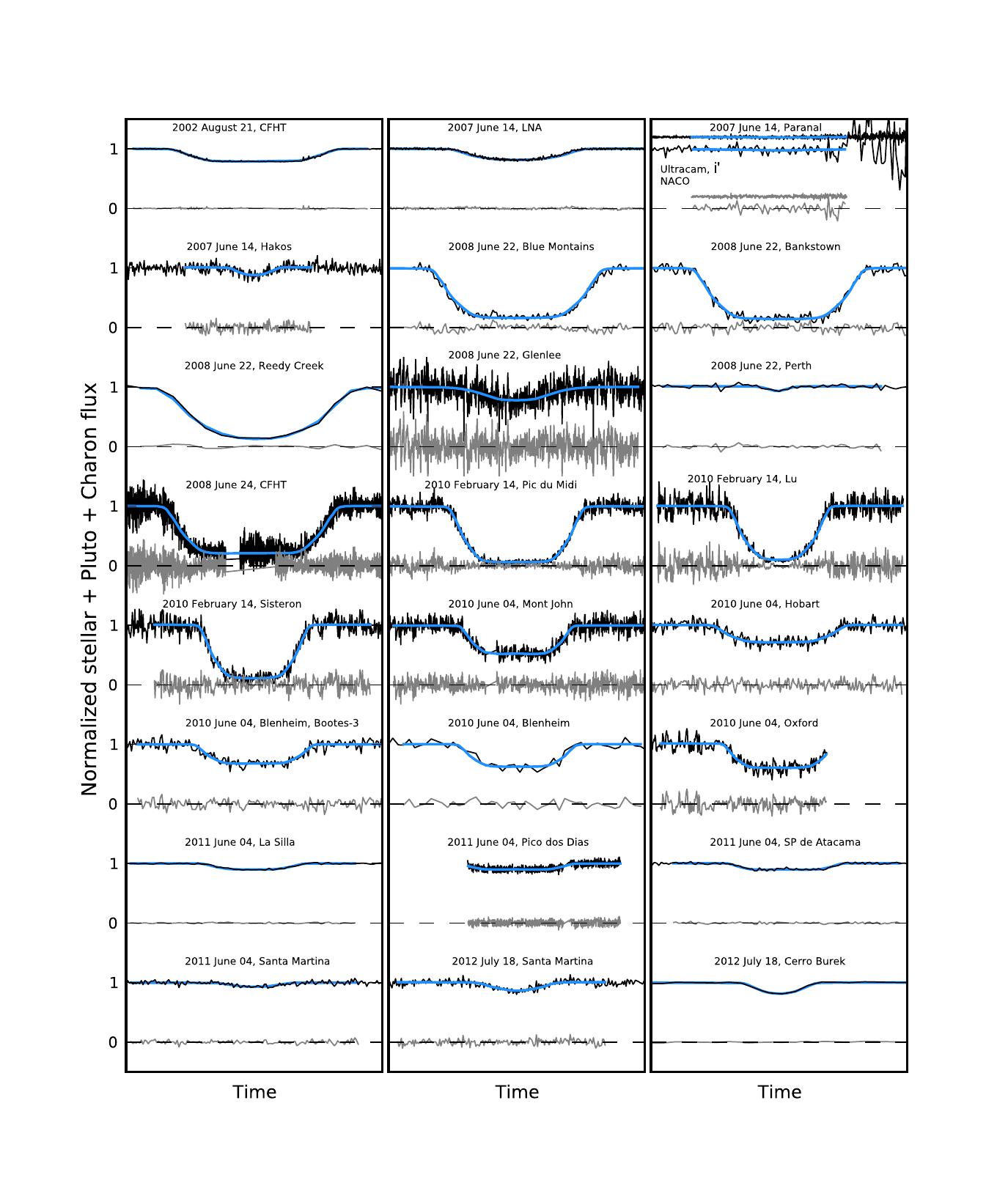}
}
\caption{%
Pluto occultation light curves obtained between 2002 and 2012. 
Blue curves are simultaneous fits (for a given date) using the DO15 temperature-radius $T(r)$ model, see text.
The residuals are plotted in gray under each light curve.
}
\label{fig_LC_fits_1}
\end{figure*}
\begin{figure*}
\centerline{%
\includegraphics[width=\paperwidth,trim=0 30 0 0,angle=0]{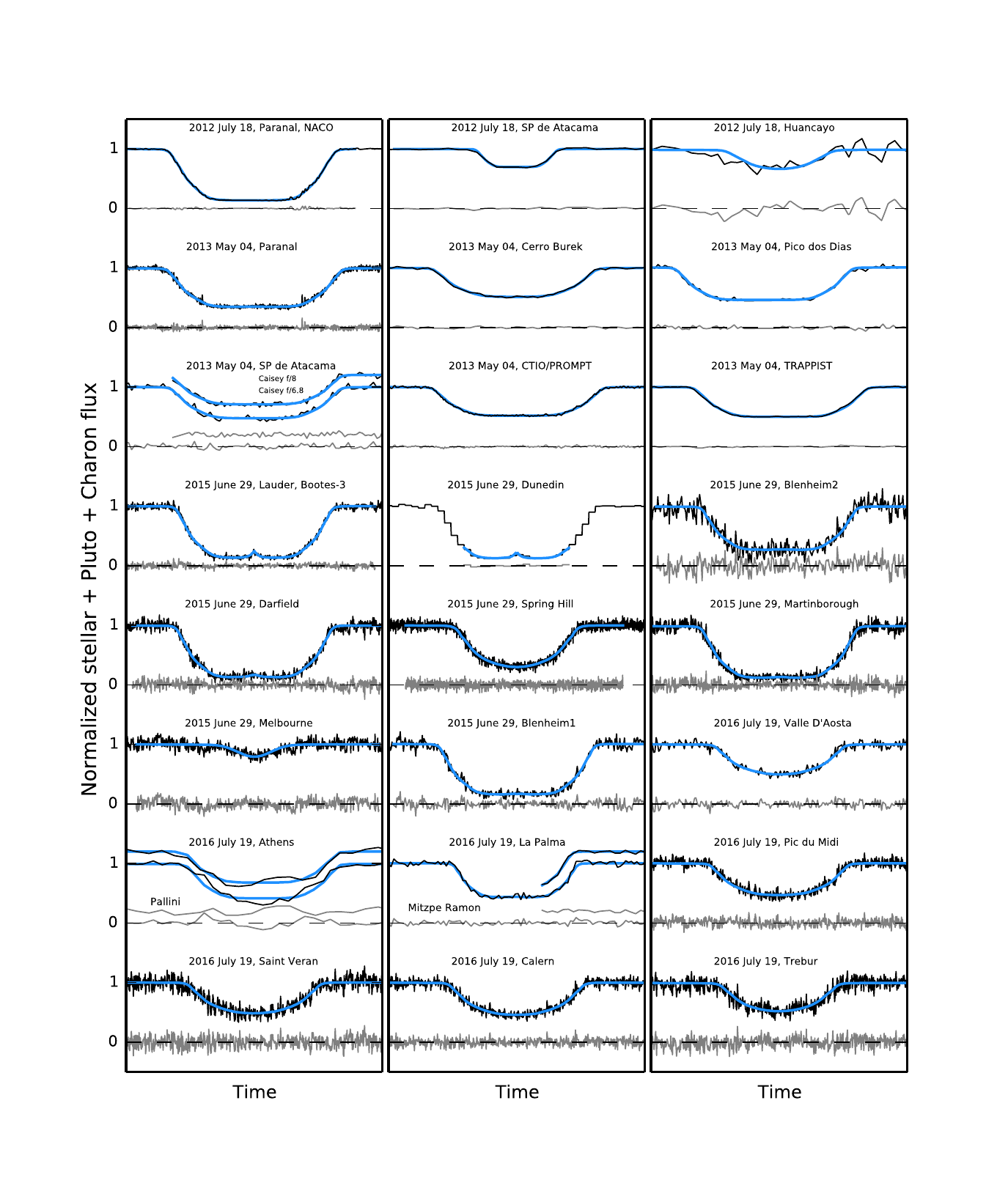}
}
\caption{%
The same as Fig.~\ref{fig_LC_fits_1} for the 2012-2016 period.
}%
\label{fig_LC_fits_2}
\end{figure*}

Taking $T(r)$ constant with time is justified by the fact that the pressure is far more sensitive to 
Pluto's surface temperature -- through the vapor pressure equilibrium equation --
than is the profile $T(r)$ to seasonal effects and heliocentric distance,
at least from a global point of view.
For instance, an increase of 1~K of the free N$_2$ ice at the surface is enough 
to multiply the equilibrium pressure by a factor of 1.7 \citep{fra09}.
Note that this is not inconsistent with our assumption that $T(r)$ is time-independent.
In fact, the overall atmospheric pressure is controlled by the temperature a few kilometers
above the surface, while our fits use a global profile $T(r)$ well above the surface.

Pluto ground-based stellar occultations probe, for the best data sets, 
altitudes from 
$\sim$5~km (pressure level $\sim$10~$\mu$bar) to
$\sim$380~km ($\sim$10~nbar level), see DO15.
Rays coming from below $\sim$5~km are detectable only near the shadow center
(typically within 50~km) where the central flash can be detected. 
The analysis is then complicated by the fact that 
double (or multiple) stellar images contribute to the flux.
Moreover, the possible presence of hazes and/or topographic features can reduce the flux, 
see Section~\ref{sec_lower_atmo}. 

Conversely, rays coming from above 380~km cause too small stellar drops ($< \sim$1\%)
to be of any use under usual ground-based observing conditions.
This said, our ray tracing method is mainly sensitive to the half-light level, 
where the star flux has been reduced by 50\%.
This currently corresponds to a radius of about 1295~km 
(or an altitude $\sim$110~km and pressure $\sim$1.6~$\mu$bar).

\subsection{Primary results}
\label{subsec_prim_results}

The ray tracing code returns the best fitting parameters, in particular 
the pressure at a prescribed radius 
(e.g. the pressure  $p_{\rm surf}$ at the surface, at radius $R_P=1187$~km) and 
Pluto's ephemeris offset perpendicular to its apparent motion, $\Delta \rho$.
The ephemeris offset along the motion is treated separately, see DO15 for details.
Error bars are obtained from the classical function 
$\chi^2 = \sum_1^N [(\phi_{\rm i,obs}-\phi_{\rm i,syn})/\sigma_i]^2$
that reflects the noise level $\sigma_i$ of each of the $N$ data points,
where $\phi_{\rm i,obs}$ and $\phi_{\rm i,syn}$ are the observed and synthetic fluxes, respectively.
An example of $\chi^{2}(\Delta \rho,p_{\rm surf})$ map is displayed in Fig.~\ref{fig_chi2_19jul16},
using a simultaneous fit to the 2015 June 29  occultation light curves.
It shows a satisfactory fit for that event, $\chi^{2}_{\rm dof} \sim$1.06.
Table~\ref{tab_pressure_time} lists the values of $\chi^{2}_{\rm dof}$ for the other occultations,
also showing satisfactory fits.
Note the slightly higher values obtained for the 2002 August 21 and 2007 June 14 events (1.52 and 1.56, respectively).
The presence of spikes in the light curve for the 2002 August 21 event (on top of the regular photometric noise)
explains this higher value, see Fig.~\ref{fig_LC_fits_1}.
From the same figure, we see that the 2007 June 14  light curves at Paranal were contaminated by clouds, 
also resulting in a slightly higher value of $\chi^{2}_{\rm dof}$.
All together, those values validate a posteriori the assumptions of 
pure N$_2$, transparent, spherical atmosphere with temperature profile constant in time.

In total, we collected and analyzed in a consistent manner 45 occultation light-curves
obtained from eleven separate ground-based stellar occultations in the interval 2002-2016 (Table~\ref{tab_sites}).
The synthetic fits to the light curves are displayed in Figs~\ref{fig_LC_fits_1} and \ref{fig_LC_fits_2}.
Fig.~\ref{fig_chord_sky} shows the occulting chords and Pluto's aspect for each event as seen from Earth. 
%

Two main consequences of those results are now discussed in turn: 
(1) the temporal evolution of Pluto's atmospheric pressure; 
(2) the structure of Pluto's lower atmosphere using the central flash of June 29, 2015.
A third product of these results 
is the update of Pluto's ephemeris using the occultation geometries between 2002 and 2016.
It will be presented in a separate paper (Desmars et al., in preparation).

\section{Pluto's atmospheric evolution}
\label{sec_pres_evolution}

\subsection{Constraints from occultations}
\begin{table}
\caption{Pluto's atmospheric pressure}
\label{tab_pressure_time}
\centering
\begin{tabular}{cccc}
\hline
\hline
 & Surface   & Pressure at  & Fit quality \\
Date   & pressure $p_{\rm surf} $& 1215 km $p_{1215}$ & $\chi^2_{\rm dof}$ \\
&       ($\mu$bar) &      ($\mu$bar) \\
\hline
\hline
1988 Jun 09 & 4.28 $\pm$ 0.44 & 2.33 $\pm$ 0.24\tablefootmark{1} & NA \\ 
2002 Aug 21 & 8.08 $\pm$ 0.18 & 4.42 $\pm$ 0.093 & 1.52 \\
2007 Jun 14 & 10.29 $\pm$ 0.44 & 5.6 $\pm$ 0.24  &  1.56 \\
2008 Jun 22 & 11.11 $\pm$ 0.59  & 6.05 $\pm$ 0.32  &  0.93 \\
2008 Jun 24 & 10.52 $\pm$ 0.51 & 5.73 $\pm$ 0.21 &  1.15 \\
2010 Feb 14 & 10.36 $\pm$ 0.4  & 5.64 $\pm$ 0.22 &  0.98 \\
2010 Jun  04 & 11.24 $\pm$ 0.96 & 6.12 $\pm$ 0.52 &  1.02 \\
2011 Jun  04 &  9.39 $\pm$ 0.70 & 5.11 $\pm$ 0.38 & 1.04 \\
2012 Jul 18 & 11.05 $\pm$ 0.08 & 6.07  $\pm$ 0.044 &  0.61 \\
2013 May 04 & 12.0 $\pm$ 0.09 & 6.53 $\pm$ 0.049 &  1.20 \\
2015 Jun 29 & 12.71 $\pm$ 0.14 & 6.92 $\pm$ 0.076 & 0.84 \\
2016 Jul 19 & 12.04 $\pm$ 0.41 & 6.61 $\pm$ 0.22 &  0.86 \\
\hline
\hline
\end{tabular}
\tablefoot{
\tablefoottext{1}{%
The value $p_{1215}$ is taken from \cite{yel97}.  
The ratio $p_{\rm surf}/p_{1215}=1.84$  of DO15's fitting model was applied to derive $p_{\rm surf}$.
Thus, the surface pressures (and their error bars) are mere scalings of the values at 1215~km.
They do \textit{not} account for systematic uncertainties caused by using an assumed 
profile (DO15 model), see discussion in subsection~\ref{subsec_transport_mod}.
The qualities of the fits (values of $\chi^2_{\rm dof}$) are commented on in subsection~\ref{subsec_prim_results}.
}
}
\end{table}					

In 2002, a ground-based stellar occultation revealed that
Pluto's atmospheric pressure had increased by a factor of 
almost two compared to its value in 1988 \citep{ell03,sic03},
although Pluto had receded from the Sun, thus globally cooling down.
In fact, models using global volatile transport did predict 
this seasonal effect, among different possible scenarios \citep{bin90,han96}.

Those models explored nitrogen cycles, 
and have been improved subsequently \citep{you12,you13,han15}.
Meanwhile, new models were developed to simulate possible scenarios for Pluto's changes 
over seasonal (248~yr) and astronomical (30 Myr) time scales,
accounting  for topography and ice viscous flow, 
as revealed by the NH flyby in July 2015 \citep{ber16,for17,ber18}.
%

The measurements obtained here provide new values of pressure vs. time, 
and are obtained using a unique light curve fitting model (taken from DO15), 
except for the 1988 occultation, see Table~\ref{tab_pressure_time}.
This model may introduce systematic biases, but it can nevertheless be used
to derive the relative evolution of pressure from date to date, and 
thus discriminates the various models of Pluto's current seasonal cycle.
In any case, the DO15 light curve fitting model appears to be close to the results derived from NH, 
see \cite{hin17} and Section~\ref{sec_lower_atmo} (Fig.~\ref{fig_n_r_rex_compa_bs}), 
so that those biases remain small.
Note that other authors also used stellar occultations 
to constrain the pressure evolution  since 1988 \citep{you08,bos15,olk15}, 
but with less comprehensive data sets. 
We do not include their results here, as they were obtained with different models
that might introduce systematic biases in the pressure values.


\subsection{Pressure evolution vs. a volatile transport model}
\label{subsec_transport_mod}

\begin{figure*}[!t]
\centerline{%
\includegraphics[totalheight=9cm, trim=0 0 0 0, angle=0]{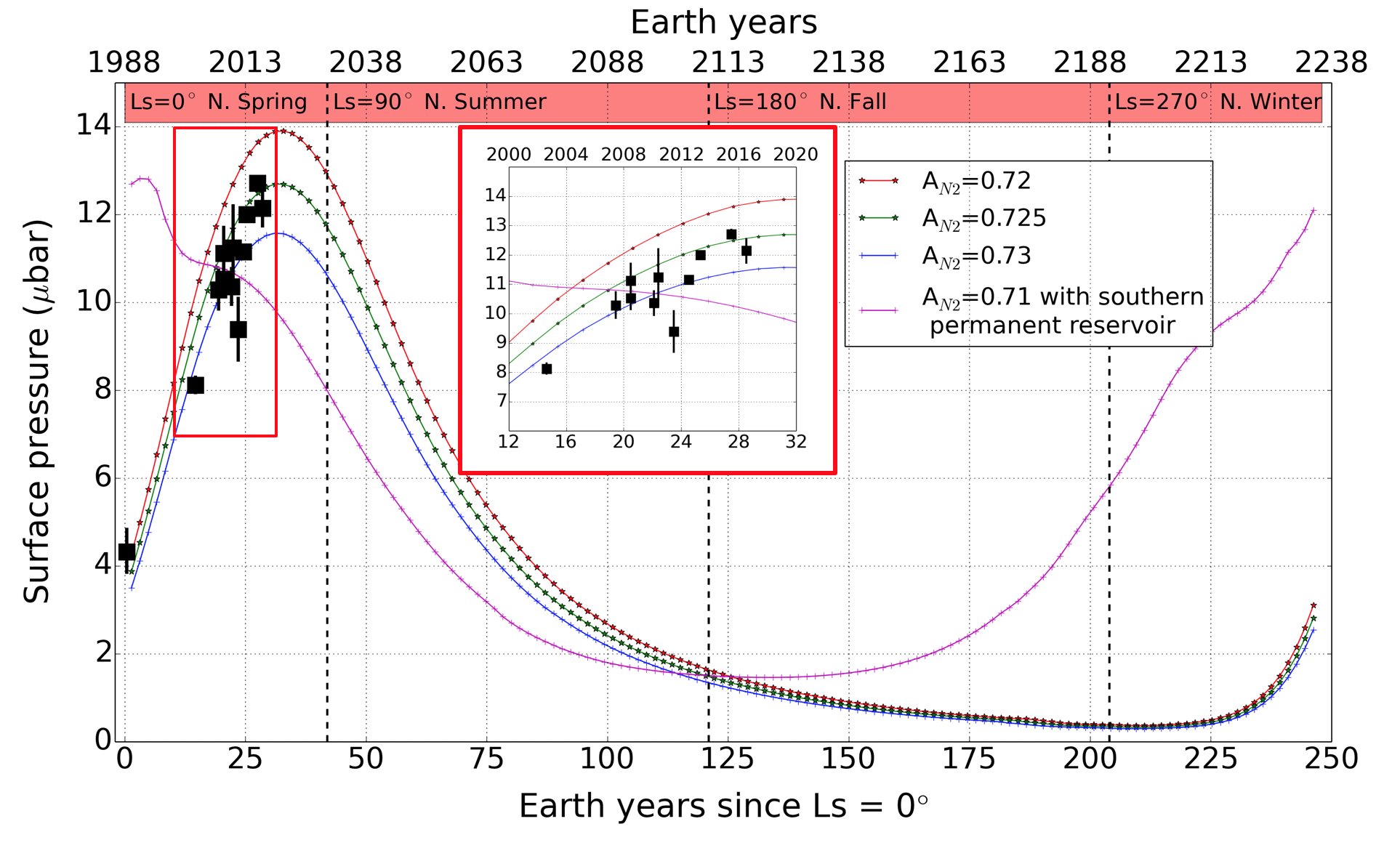}
}
\caption[Pluto's atmospheric evolution interpretation]{
Typical modeled annual evolution of surface pressure obtained with LMD Pluto volatile transport model, 
assuming permanent deposits of N$_2$ ice inside Sputnik Planitia and in the depression of mid-northern latitudes, 
a uniform soil seasonal thermal inertia of 800 J s$^{-1/2}$ m$^{-2}$ K$^{-1}$,
an emissivity $\epsilon_{\rm N2}$ = 0.8 and albedo range $A_{\rm N2}$ = 0.72-0.73 for N$_2$ ice, 
chosen to yield a surface pressure near 10-11 $\mu$bar in July 2015. 
The black dots with error bars show the surface pressure ($p_{\rm surf}$) inferred from stellar occultation 
pressure measurements (see Table~\ref{tab_pressure_time}). 
The curve in magenta corresponds to a similar simulation but assuming a permanent N$_2$ ice reservoir 
in the south hemisphere between 52.5$\degr$ and 67.5$\degr$ S, which leads to a pressure peak in 1990.
}
\label{fig_pressure_time}
\end{figure*}


Table~\ref{tab_pressure_time} provides the pressure derived at each date,
at the reference radius $r=1215$~km (altitude 28~km), 
their scaled values at the surface using the DO15 model,
as well as the pressure previously derived from the 1988 June 09 occultation.
Figure~\ref{fig_pressure_time} displays the resulting  pressure evolution during the time span 1988-2016.
As discussed in the previous subsection, 
even if the use of the DO15 model induces biases on $p_{\rm surf}$, 
it should be a good proxy for the global evolution of the atmosphere, and as such, 
provides relevant constrain for Pluto's seasonal models.

We interpret our occultation results in the frame of the Pluto volatile transport model 
developed at the Laboratoire de M\'et\'eorologie Dynamique (LMD).
It is designed to simulate the volatile cycles over seasonal and 
astronomical times scales on the whole planetary sphere \citep{ber16,for17,ber18}.
We use the latest, most realistic, version of the model featuring the topography map of Pluto \citep{sch18a} 
and large ice reservoirs \citep{ber18}. 
In particular, we place permanent reservoirs of nitrogen ice in the Sputnik Planitia basin and
in the depressions at mid-northern latitudes (30$^\circ$N, 60$^\circ$N), 
as detected by NH \citep{smt17} and modeled in \cite{ber18}.

Fig.~\ref{fig_pressure_time} shows the annual evolution of surface pressure obtained with the model,
compared to the data. 
This evolution is consistent with the continuous increase of pressure observed 
since equinox in 1988, reaching an overall factor of almost three in 2016. 
This results from the progressive heating of the nitrogen ice in Sputnik Planitia and in the northern mid-latitudes, 
when those areas were exposed to the Sun just after the northern spring equinox in 1988,
and close in time to the perihelion of 1989, as detailed in \cite{ber16}.

The model predicts that the pressure will reach its peak value and then drop in the next few years, due to: 
  
(1) the orbitally-driven decline of insolation over Sputnik Planitia and the northern mid-latitude deposits; 
  
(2) the fact that nitrogen condenses more intensely in the colder southern part of Sputnik Planitia, 
thus precipitating and hastening the pressure drop. 

The climate model has several free parameters:
the distribution of nitrogen ice, its Bond albedo and emissivity and
the thermal inertia of the subsurface (soil).
However, the large number of observation points and the recent NH observations provide
strong constraints for those parameters,
leading to an almost unique solution. 

First, our observations restrict the possible N$_2$ ice surface distribution. 
Indeed, the southern hemisphere of Pluto is not expected to be significantly covered 
by nitrogen ice at the present time, because otherwise the peak of surface pressure 
would have occurred much earlier than 2015, as suggested by the model simulations 
(Fig.~\ref{fig_pressure_time}).
With our model, we obtain a peak of pressure after 2015 only when considering little 
mid-latitudinal nitrogen deposits (or no deposit at all) in the southern hemisphere.

In our simulation, nitrogen does not condense much in the polar night (outside Sputnik Planitia), 
in spite of the length of the southern fall and winter. 
This is because in Pluto conditions, depending of the subsurface thermal inertia,  
the heat stored in the southern hemisphere during the previous southern hemisphere summer can keep the surface temperature 
above the nitrogen frost point throughout the cold season, or 
at least strongly limit the nitrogen condensation. 

Consequently, the data points provide us with a second constraint,
which is a relatively high subsurface thermal inertia 
so that nitrogen does not condense much in the southern polar night. 
Using a thermal inertia between 700-900 J s$^{-1/2}$ m$^{-2}$ K$^{-1}$ 
permits us to obtain a surface pressure ratio 
$(p_{\rm surf,2015}/p_{\rm surf,1988})$ of  around 2.5-3, as observed. 
Higher (resp. lower) thermal inertia tend to lower (resp. increase) this ratio,
as shown in Fig.~(2a) of \citet{ber16}. 

Finally, the nitrogen cycle is very sensitive to the nitrogen ice Bond albedo $A_{N2}$ and emissivity $\epsilon_{N2}$,
and only a small range for these parameters allows for a satisfactory match to the observations.
%
%
Fig.~\ref{fig_pressure_time} illustrates that point. 
To understand it, one can do the thought experiment of imagining 
Pluto with a flat and isothermal surface at vapor pressure equilibrium. 
A rough estimate of the equilibrium temperature is provided by the classical equation:
\begin{equation*}
\epsilon_{N2} \sigma T^{4}= \left (1-A_{N2} \right)\frac{F}{4},
\end{equation*}
where $F$ is the solar constant at Pluto and 
$\sigma = 5.67 \times 10^{-8}$ W m$^{-2}$ K$^{4}$ is the Stefan-Boltzmann constant. 
The surface pressure $p_{\rm surf}$  is then estimated from the surface temperature $T_{\rm surf}$ 
assuming N$_2$  vapor pressure equilibrium \citep{fra09}.
Consequently, the surface pressure data set inferred from stellar occultations provide us 
with a constraint on $(1-A_{N2})/\epsilon_{N2}$. 
In practice, in the model, we assume large grains for N$_2$ ice and we fix the emissivity 
at a relatively high value $\epsilon_{N2}=0.8$ \citep{lel11}.
Taking $F=1.26$ W m$^{-2}$ (in 2015) and assuming $A_{N2}$ = 0.72, 
we find $T_{\rm surf} = 37.3$~K, and a corresponding vapor pressure
$p_{\rm surf}=$ 14.8~$\mu$bar for the N$_2$ ice at the surface.
With $A_{N2} = 0.73$, we obtain $T_{\rm surf}=37.0$~K and $p_{\rm surf}=12.0$~$\mu$bar.
Thus, the simple equation above provides pressure values that 
are consistent with the volatile transport model displayed in Fig.~\ref{fig_pressure_time}.
It then can be used to show that
decreasing the nitrogen ice albedo by only 0.01 
leads to an increase of surface pressure in 2015 by a large amount of 25\%.

\begin{figure}
\centering
\includegraphics[width=90mm,trim=0 0 0 0]{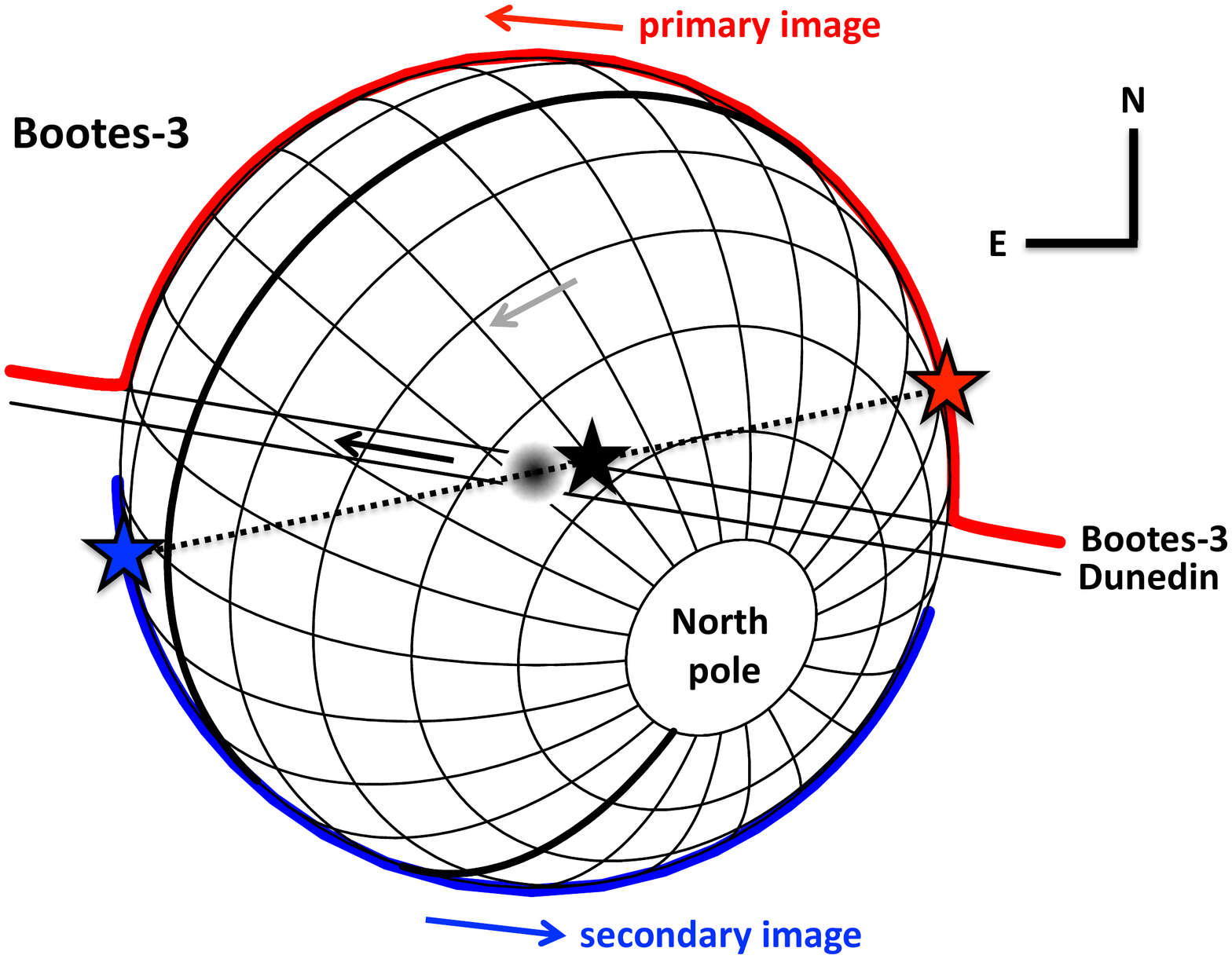}
\includegraphics[width=90mm,trim=0 0 0 0]{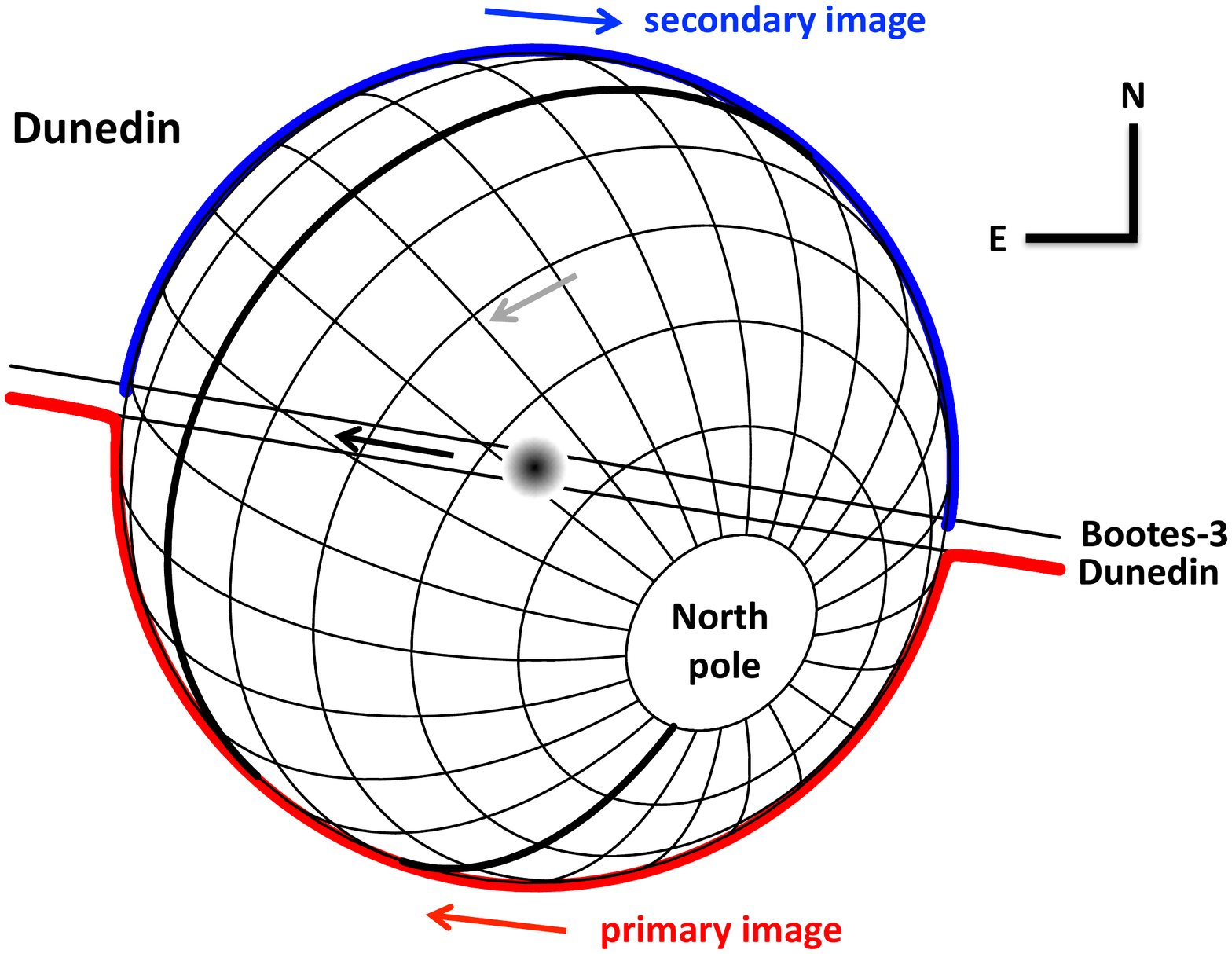}
\caption{%
The reconstructed geometry of the June 29, 2015 Pluto stellar occultation.
Celestial north is at top and celestial east at left, see labels N and E.
%
%
The equator and prime meridian (facing Charon) are drawn as thicker lines.
The direction of Pluto's rotation is along the gray arrow.
In the two panels, 
the stellar motion relative to Pluto is shown as black solid lines as seen from the Bootes-3 and Dunedin stations,
with direction of motion marked by the black arrow.
%
%
The shaded region at center roughly indicates the zone where a central flash could be detected.
In the upper panel, the red and blue lines are the trajectories of 
the primary and secondary stellar images, respectively, as seen from Bootes-3.
Lower panel: the same for the stellar images as seen from Dunedin.
%
%
For a spherical atmosphere, 
the position of the star in the sky plane, 
the center of Pluto
and the two images are aligned, 
as shown in the upper panel (see the dotted line connecting the star symbols).
%
}%
\label{fig_chords_boo_dun}
\end{figure}

\section{Pluto's lower atmosphere}
\label{sec_lower_atmo}

\subsection{The June 29, 2015 occultation}

The June 29, 2015 event provided seven chords across Pluto's atmosphere, 
see Table~\ref{tab_sites} and Fig.~\ref{fig_chord_sky}.
A first analysis of this event is presented in \cite{sic16}.
The two southernmost stations (Bootes-3 and Dunedin) probed the central flash region
(Fig.~\ref{fig_chords_boo_dun}).
This was a unique opportunity to study Pluto's lower atmosphere
a mere fortnight before the NH flyby (July 14, 2015).
During this short time lapse, 
we may assume that the atmosphere did not suffer significant global changes. 

\begin{figure*}
\centering
\includegraphics[width=90mm,trim=0 0 0 0,angle=0]{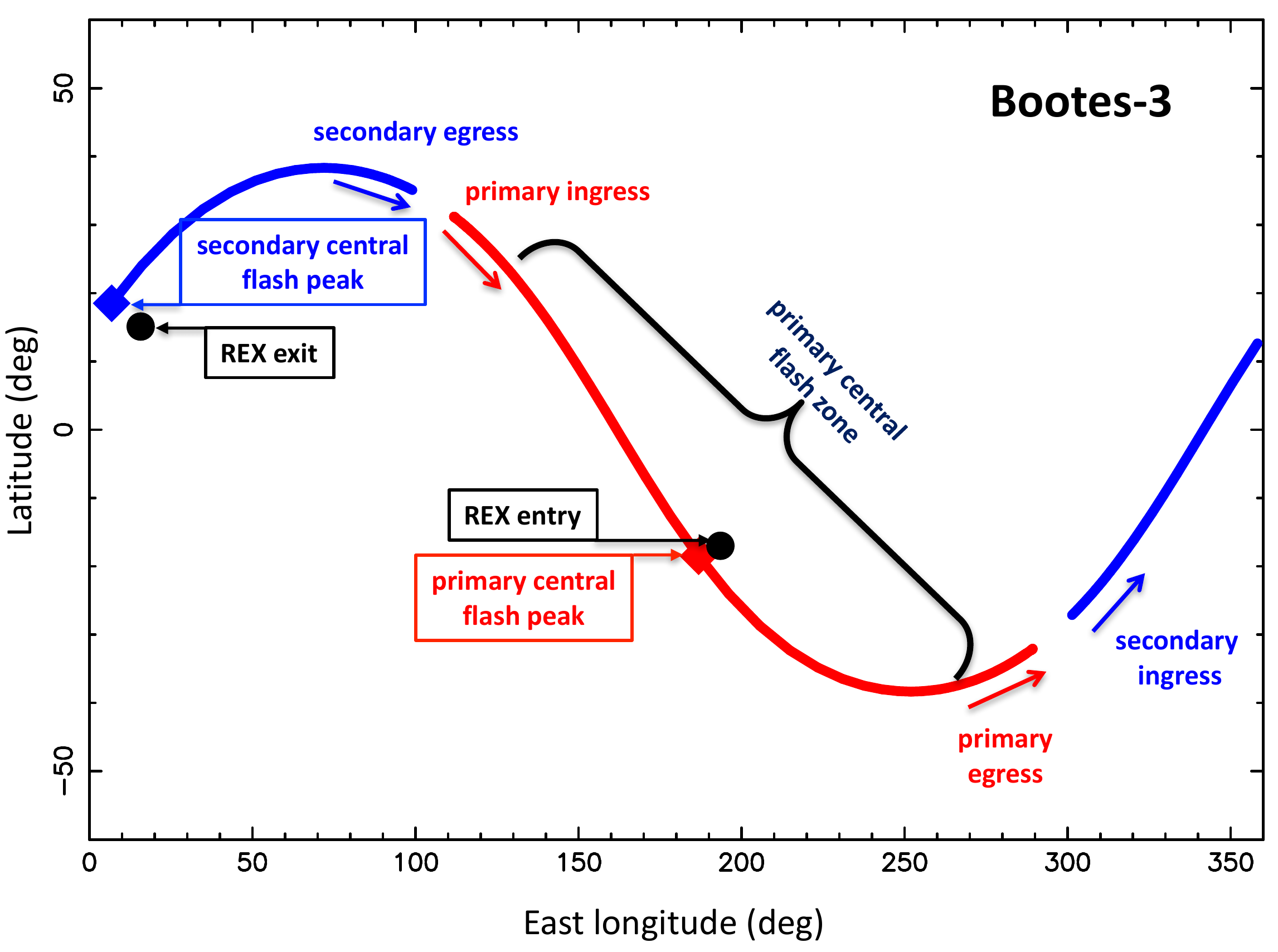}
\includegraphics[width=90mm,trim=0 0 0 0,angle=0]{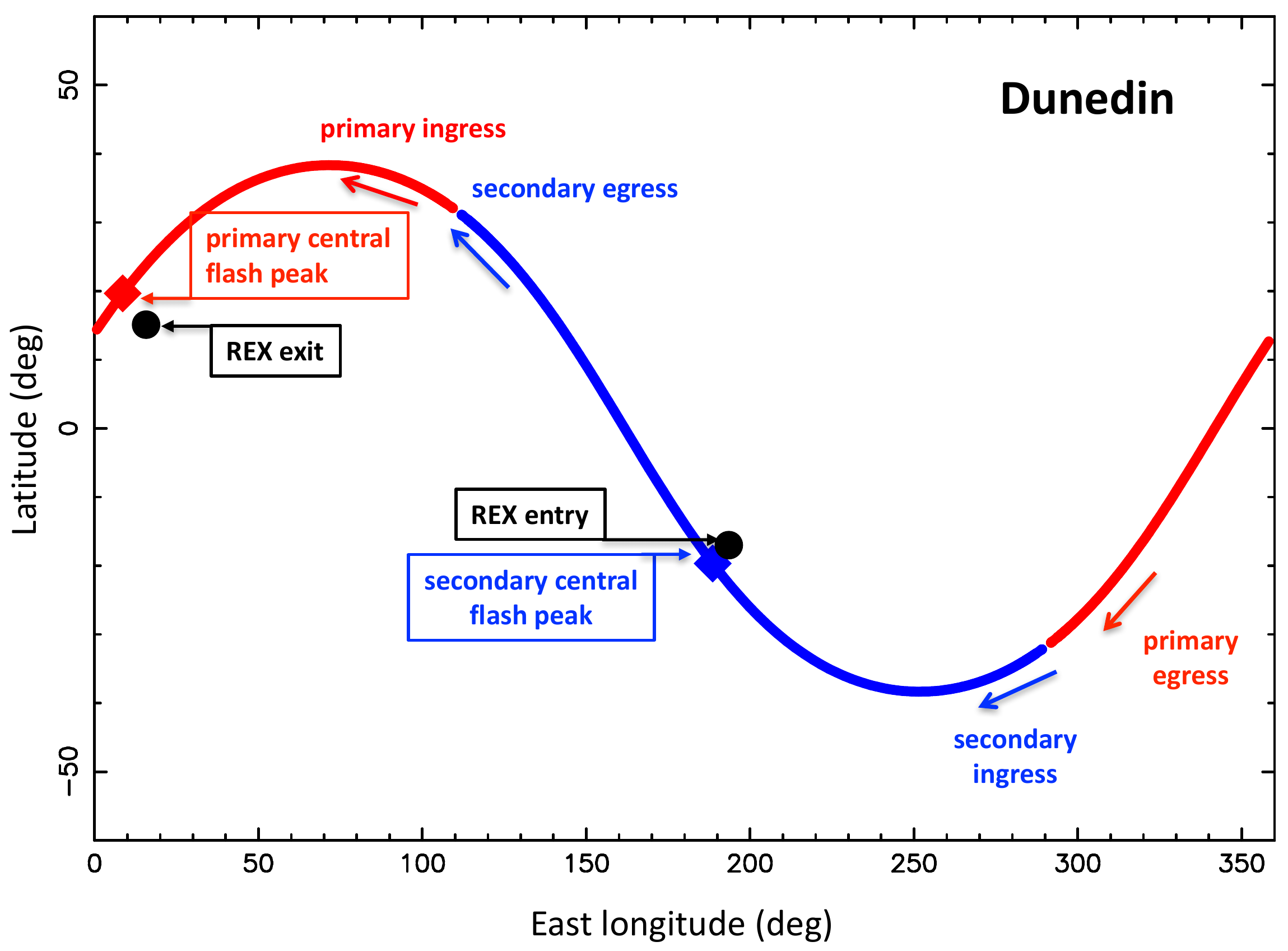}
\caption{%
Left panel~- 
Traces of the primary (red) and secondary (blue) stellar images observed 
at Bootes-3, as deduced from Fig.~\ref{fig_chords_boo_dun}.
The arrows indicate the direction of motion.
``Ingress" (resp. ``egress") refers to the disappearance (resp. re-appearance) 
of the images into Pluto's atmosphere.
The diamond-shaped symbols mark the positions of the image at the peak of the flash,
corresponding to the time of closest approach of the respective station to the shadow center. 
In total, the primary image scanned longitudes from 120$\degr$ to 270$\degr$,
while the secondary image scanned longitudes from 310$\degr$ to 360$\degr$ and then from 0 to 70$\degr$.
The brace indicates the total duration of the primary flash 
($\sim$15 s, see Fig.~\ref{fig_fit_rex_boo_dun}) at Bootes-3,
covering a rather large region of more than 120$\degr$ in longitude.
A similar extension applies to the secondary flash,
but the brace has not been drawn for sake of clarity.
The black bullets are the locations of the REX measurements at entry and exit \citep{hin17}. 
Note the casual proximity of the REX points and the June 29, 2015 flash peaks.
Right panel~- The same for the Dunedin station, where the brace has not been repeated. 
Note that the tracks and motions of the primary and secondary images 
are essentially swapped between the two stations.
}
\label{fig_lon_lat_boot_dun}
\end{figure*}

For a spherical atmosphere, there are at any moment two stellar images,
a primary (near limb) image and a secondary (far limb) image that are aligned
with Pluto's center and the star position, as projected in the sky plane,
see Fig.~\ref{fig_chords_boo_dun}.
Since the ray tracing code provides the refraction angle corresponding to each image, 
their positions along Pluto's limb can be determined at any time (Fig.~\ref{fig_chords_boo_dun}),
and then projected onto Pluto's surface (Fig.~\ref{fig_lon_lat_boot_dun}).

\subsection{Comparison with the REX results}

The REX instrument recorded an uplinked 
4.2~cm radio signal sent from Earth.
The phase shift due to the neutral atmosphere was then used to retrieve the 
$n(r)$, $p(r)$ and $T(r)$ profiles through an inversion method 
and the usual ideal gas and hydrostatic assumptions \citep{hin17}.
The REX radio occultation probed two opposite points of Pluto as the 
signal disappeared behind the limb (entry) and re-appeared (exit),
see Fig.~\ref{fig_lon_lat_boot_dun}. 
Note that the REX entry point is at the southeast margin of Sputnik Planitia,
a depression that is typically 4~km below the surrounding terrains, 
see \cite{hin17} for details. 

\begin{table*}
\caption{Regions probed by the central flash (June 29, 2015) and REX experiment (July 14, 2015)}
\label{tab_flash_29jun15}
\centering
\begin{tabular}{lccc}
\hline
 & Time (UT)\tablefootmark{1}  & Location on surface & Local solar time\tablefootmark{2} \\
\hline
\multicolumn{4}{c}{June 29, 2015} \\
\hline
Bootes-3, primary image     &  16:52:54.8  & 186.8$^\circ$E,  18.5$^\circ$S & 7.67 (sunrise) \\
\hline
Bootes-3, secondary image & 16:52:54.8  &     6.8$^\circ$E,  18.5$^\circ$N & 19.67 (sunset) \\
\hline
Dunedin, primary image      &  16:52:56.0  &    8.6$^\circ$E,  19.7$^\circ$N  & 19.79 (sunset) \\
\hline
Dunedin, secondary image & 16:52:56.0   &  188.6$^\circ$E, 19.7$^\circ$S  & 7.79 (sunrise) \\
\hline
\multicolumn{4}{c}{NH radio experiment (REX), July 14, 2015} \\
\hline
entry &  12:45:15.4 & 193.5$^\circ$E, 17.0$^\circ$S & 16.52 (sunset) \\
\hline
exit &  12:56:29.0 & 15.7$^\circ$E, 15.1$^\circ$N & 4.70 (sunrise) \\
\hline
\end{tabular}
%
\tablefoot{%
\tablefoottext{1}{For the ground-based observations, this is the time of closest approach to shadow center
\citep{sic16}, for the REX experiment, this the beginning and end of occultation by the solid body \citep{hin17}.}
\tablefoottext{2}{One ``hour" corresponds to a rotation of Pluto of 15$^\circ$.
A local time before (resp. after) 12.0~h means morning (resp. evening) limb.}
}
\end{table*} 

Note also the (serendipitous) proximity of 
the regions scanned by the June 29, 2015 central flash and 
the two zones probed by REX at entry and exit. 
This permits relevant tests of the REX profiles against the central flash structure.
The local circumstances on Pluto for the central flash and the REX occultation are
summarized in Table~\ref{tab_flash_29jun15}. 
However, that the local times are swapped between our observations and REX suboccultation points:
the sunrise regions of one being the sunset places of the other, and vice versa,
see the discussion below.

The REX profiles are in good general agreement with those derived by \cite{sic16} 
-- based itself on the DO15 procedure --
between the altitudes of 5~km and 115~km 
(Figs.~\ref{fig_n_r_rex_compa_bs} and \ref{fig_T_r_rex_compa_bs}), 
thus validating our approach. 
However, we see discrepancies at altitudes below $\sim$25 km ($r < 1212$~km), 
in the region where the REX entry and exit profiles diverge from one another.

Part of those differences may stem from the swapping of the sunrise and sunset limbs 
between the REX measurements and our observations, 
and to the fact that a diurnal sublimation/condensation cycle of N$_2$ occurs over Sputnik Planitia.
Then, lower temperatures just above the surface are expected at the end of the afternoon in that region, 
after an entire  day of sublimation \citep{hin17}. 
Conversely, a warmer profile could prevail at sunrise, after an entire night of condensation.
This warmer profile would then be more in agreement with the DO15 temperature profile.

However, the difference between the REX (red) and DO15 (black) profiles in Fig.~\ref{fig_T_r_rex_compa_bs}
remains large (more than 20~K at a given radius). 
This is much larger than expected from current GCMs (e.g. \citealt{for17}, Fig.~7), 
which predict diurnal variations of less than 5~K at altitude levels 1-2~km above Sputnik Planitia, 
and less than 1~K in the $\sim$4-7~km region that causes the flash \citep{sic16}.
In practice, \citealt{for17} predict that above 5-km,
the temperature should be uniform over the entire planet at a given radius.
This is in contrast to REX observations, 
that reveal different temperature profiles below 25~km (Fig.~\ref{fig_T_r_rex_compa_bs}).
Thus, ingredients are still missing to fully understand REX observations,
for instance the radiative impact of organic hazes, 
an issue that remains out of the scope of this paper.

Note that the entry REX profile goes deeper than the exit profile.
This reflects the fact that the nominal Pluto's radii are at
$1187.4 \pm 3.6$~km at entry and $1192.4 \pm 3.6$~km at exit \citep{hin17}.
This discrepancy is not significant considering the uncertainties on each radius. 
However, the examination of Fig.~\ref{fig_p_r_rex_compa_bs_zoom} 
shows that the most probable explanation
of this mismatch is that REX probed higher terrains at exit than at entry, 
then providing the same pressure at a given planetocentric radius.
This is the hypothesis that we will adopt here, 
which is furthermore supported by the fact that the REX entry point is actually near
the depressed region Sputnik Planitia.
More precisely, the REX solution for the radius at entry ($1187.4 \pm 3.6$~km)
is fully consistent with the radius derived from NH stereo images at the same location, 
$1186.5 \pm 1.6$~km \citep{hin17}.
This said, note that our data do not have enough sensitivity to constrain
the absolute vertical scale of the density profiles at a better level than the REX solution
($\pm 3.6$~km), see next subsection.

\subsection{The June 29, 2015 central flash}

The REX profiles extend from the surface
(with pressures of $12.8 \pm 0.7$ and $10.2 \pm 0.7$~$\mu$bar at entry and exit, respectively) 
up to about 115~km, where the pressure drops to $\sim$1.2~$\mu$bar. 
Meanwhile, \cite{sic16} derive a consistent surface pressure of 12.7~$\mu$bar,
with error domains that are discussed later.

This said, the DO15-type thermal profile for the stratosphere (also called inversion layer) that extends between
the surface and the temperature maximum at $r=1215$~km is assumed to have a hyperbolic shape. 
The DO15 profile stops at its bottom at the point where it crosses the vapor pressure equilibrium line, 
thus defining the surface (assuming no troposphere).
While the adopted functional form captures the gross structure of the thermal profile, it remains arbitrary.
In fact, as the error bars of the REX profiles decrease with decreasing altitude, 
it becomes clear that the DO15 profile overestimates the temperature by tens of degrees (compared to REX)
in the stratosphere as one approaches the surface.
Also, it ends up at the surface with a thermal gradient (16~K~km$^{-1}$, see Fig.~\ref{fig_T_r_rex_compa_bs}) 
that is much stronger than in the REX profiles,
where it is always less that 10~K~km$^{-1}$ in the stratosphere.
As discussed in the previous subsection, however, the N$_2$ diurnal cycle might induce 
a warmer temperature profile (after nighttime condensation) at a few km altitude above Sputnik Planitia.
This would result in a larger thermal gradient that would be closer to the DO15 profile,
but still too far away from it according to GCM models, as discussed previously.

\begin{figure}
\centering
\includegraphics[width=90mm,trim=0 0 0 0,angle=0]{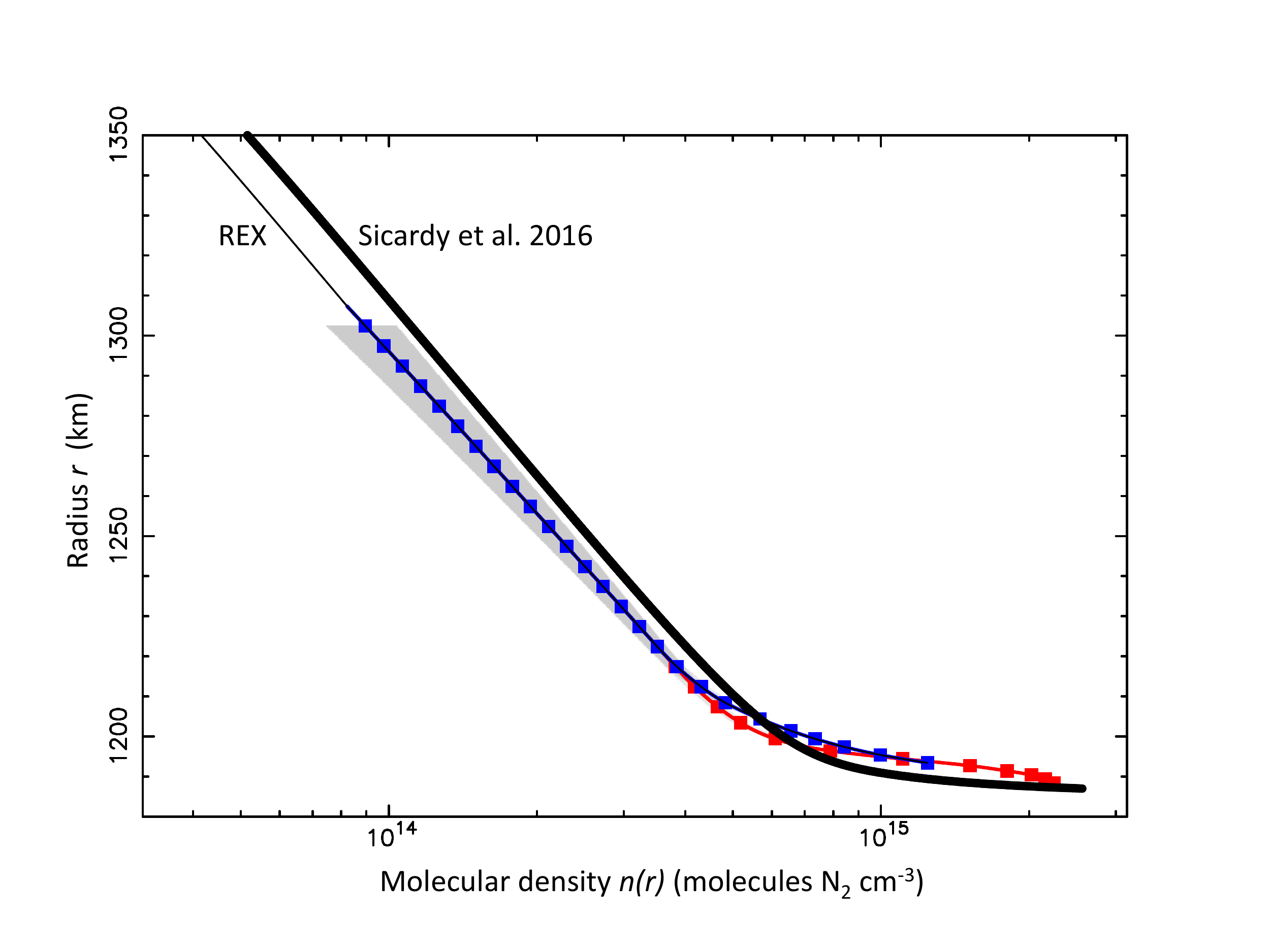}
\caption{%
Red and blue squares: 
the REX radio occultation N$_2$ density profiles,
with the shaded area indicating the 1-$\sigma$ error bar domain \citep{hin17}.
Below 1220~km, the errors decrease and become unnoticeable in this plot. 
The entry (resp. exit) profile is given from $r=1188.4$~km (resp. 1193.4~km), 
up to 1302.4~km, where the error bars become too large for a reliable profile to be retrieved.
Note that by construction, the REX entry and exit profiles are \it identical \rm for $r > 1220$~km.
Below that radius, the two profiles diverge significantly,
due to different physical conditions of the boundary layer just above the surface (Fig.~\ref{fig_T_r_rex_compa_bs}).
The solid red and blue lines connecting the squares are spline interpolations
of the REX profiles that are used in our ray tracing code, see text.
The REX profile is extended above $r=1302.4$~km as a thin solid line,
by adopting a scaled version of the June 29, 2015 profile 
(i.e. a mere translation of the thick solid line in this $(\log_{10}(n),r)$ plot), 
while ensuring continuity with the REX profile. 
Thick solid line: the profile derived by \cite{sic16} using the DO15 light curve fitting model.
The formal 1-$\sigma$ error bar of this profile is smaller than the thickness of the line,
but does not account for possible biases, see text.
}
\label{fig_n_r_rex_compa_bs}
\end{figure}

\begin{figure}
\centering
\includegraphics[width=90mm,trim=0 0 0 0,angle=0]{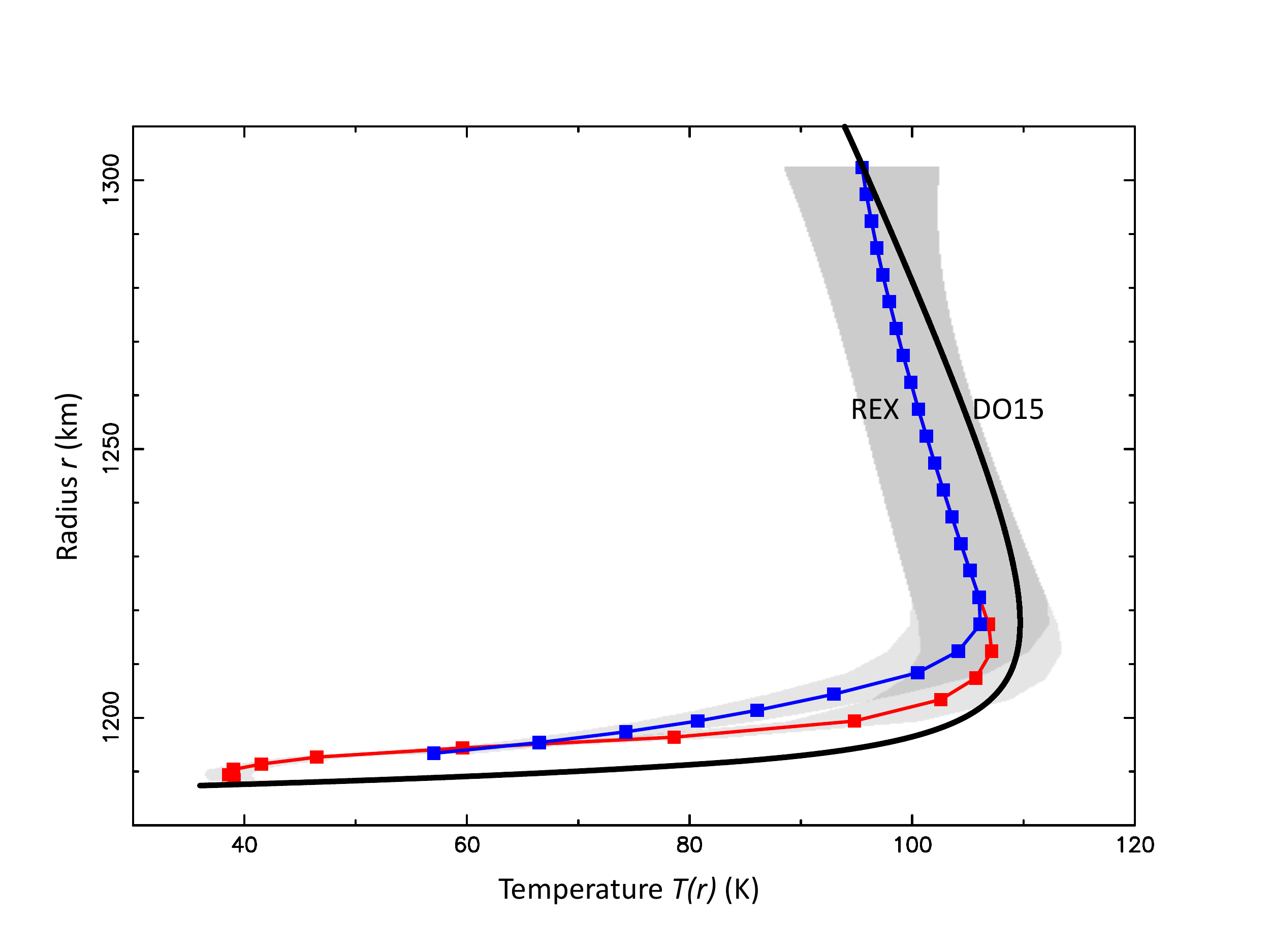}
\caption{%
The same as in Fig.~\ref{fig_n_r_rex_compa_bs} for the temperature profiles $T(r)$. 
%
By construction, 
the REX profile uses a boundary condition $T_b= 95.5$~K at the reference radius $r_b= 1302.4$~km,
in order to connect it to the DO15 profile (solid black line).
Thus, the intersection of the REX and DO15 profiles at $r_b$ is  a
mere result of the choice of $T_b$, not a measurement.
There is no formal error bars on the \citealt{sic16}'s temperature profile, 
as most of the errors come in this case from biases,  see text.
}
\label{fig_T_r_rex_compa_bs}
\end{figure}

\begin{figure}
\centering
\includegraphics[width=90mm,trim=0 0 0 0,angle=0]{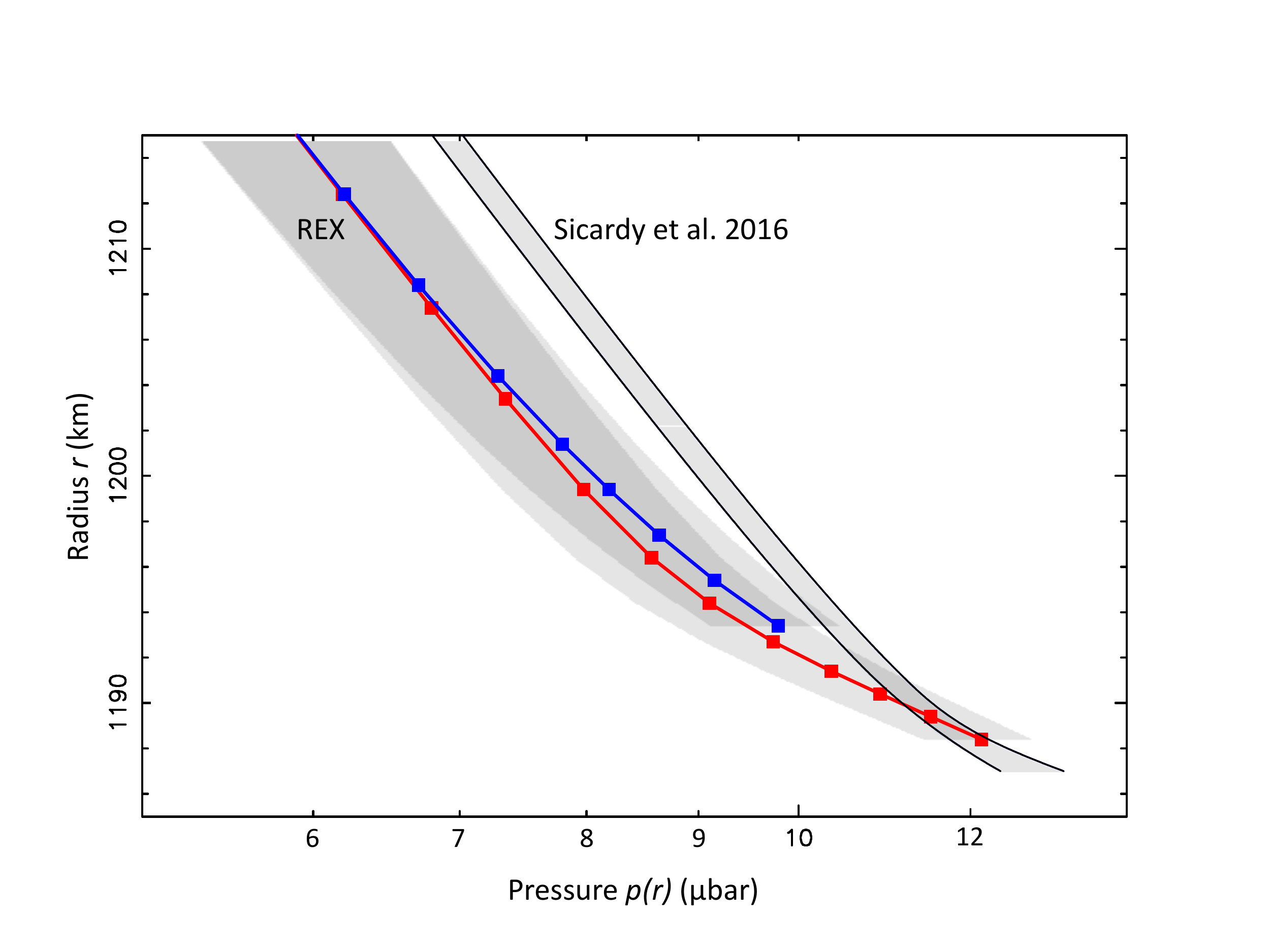}
\caption{%
The same as in Fig.~\ref{fig_n_r_rex_compa_bs}, 
but for the pressure profiles $p(r)$. 
The gray region encompassing the \citealt{sic16}'s profile and delimited by thin solid lines
is the uncertainty domain discussed by those authors.
}%
\label{fig_p_r_rex_compa_bs_zoom}
\end{figure}

In that context, we have tested the REX profiles after modifying our ray tracing procedure 
to generate new synthetic central flashes. 
We now account for the fact that the two stellar images that travel along Pluto's limb probe different density profiles.
To simplify as much as possible the problem, 
we assume that the stellar images that follow the northern and southern limbs probe an atmosphere that, 
respectively, has the entry and exit REX density profiles, 
in conformity with the geometry described in Fig.~\ref{fig_lon_lat_boot_dun}.
This is an oversimplified approach as the stellar images actually scan rather large portions of the limb,
not just the REX entry and exit points (Fig.~\ref{fig_lon_lat_boot_dun}). 
However, this exercise allows us to assess how different density profiles may affect the shape of the central flash.
To ensure smooth synthetic profiles, the discrete REX points have been interpolated by spline functions, 
using a vertical sampling of 25~meters. 
Finally, above the radius $r=1302.4$~km, the REX profiles have been extrapolated 
using a scaled version of the DO15 profile (see details in Fig.~\ref{fig_n_r_rex_compa_bs}).

\begin{figure*}

\centerline{\includegraphics[totalheight=120mm,trim=0 0 0 50,angle=0]{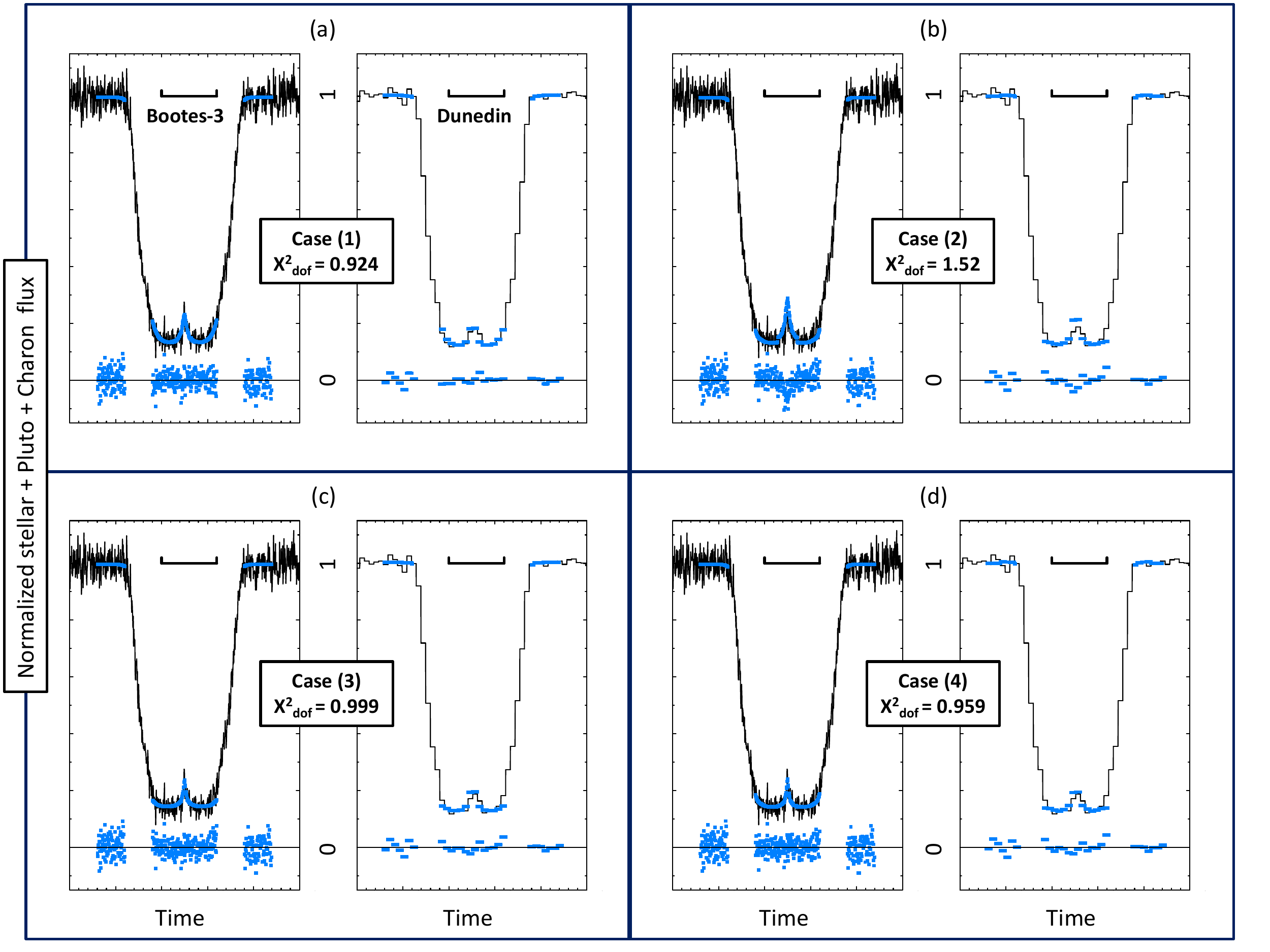}}

\caption{%
In each panel, the synthetic fits to the Bootes-3 (left) and Dunedin (right) observations 
of June 29, 2015 are shown as blue points,
together with the residuals (observations minus model) under each light curve, 
for each of the cases discussed in the text.
The tick marks on the time axis are plotted every 10~s, 
and the horizontal bars above each curve show the one-minute interval from 16h 52m 30 to 16h 53m 30s UT.
(a)
The best fits to the Bootes-3 and Dunedin light curves using the DO15 light curve fitting model \citep{sic16}, 
see also Figs.~\ref{fig_n_r_rex_compa_bs}-\ref{fig_T_r_rex_compa_bs}.
(b)
The same but using the nominal REX density profile. 
Note that the synthetic flashes are too high at both stations.
(c)
The same, after multiplying the REX density profiles by a factor $f=0.805$ and moving 
Pluto's shadow 17~km north of the solution of \cite{sic16}.
(d)
The same using the nominal REX profiles, but with a topographic feature of height $h=1.35$~km
that blocks the stellar image during part of its motion along the southern Pluto limb (Fig.~\ref{fig_chords_boo_dun}). 
Pluto's shadow has now been moved by 19.5~km north of the solution of \cite{sic16}.
In each panel, 
the value of the $\chi^2$ function per degree of freedom ($\chi^2_{\rm dof}$)
provides an estimation of the quality of the fit, 
see text for discussion. 
}%
\label{fig_fit_rex_boo_dun}
\end{figure*}

Because we want to test the shape of the central flash only, 
we restrict the generation of the synthetic light curves to the bottom parts of the occultation.
We also include in the fit two intervals that bracket the event outside the occultation, 
where we know that the flux must be unity (Fig.~\ref{fig_fit_rex_boo_dun}).
Those external parts do not discriminate the various models, 
but serve to scale properly the general stellar drop.  
Thus, the steep descents and ascents of the occultation light curves are avoided,
as they would provide too much weight to the fits.
Finally, since no calibrations of the light curves are available to assess 
Pluto's contribution $\phi_P$ to the observed flux,
a linear least-square fit of the synthetic flux to the data has been performed before calculating the residuals.
This introduces a supplementary adjustable parameter, $\phi_P$ to the fits.

Four simple scenarios are considered.
(1) We first use the original model of \cite{sic16} to generate the light curves.
(2) We take the REX density profiles at face value and use the modified ray tracing model described above,
fixing Pluto's ephemeris offset as determined in Case (1). 
(3) We apply an adjustable, uniform scaling factor $f$ to the two REX density profiles
(which thus also applies to the pressure profile since the temperature is fixed),
and we adjust Pluto's ephemeris offset accordingly.
(4) Turning back to the REX density profiles of Case (2), 
we assume that a topographic feature of height $h$ (on top of the REX exit radius, 1192.4~km) blocks 
the stellar image generated by the REX exit profile, 
i.e. that the stellar image that travels along the southern limb (Fig.~\ref{fig_chords_boo_dun})
is turned off below a planetocentric radius $1192.4+h$~km.

It should be noted that the amplitude of the synthetic flash is insensitive
to the absolute altitude scale that we use for the REX density profiles, 
to within the  $\pm 3.6$~km uncertainty discussed in the previous subsection.
For instance, displacing the REX entry profile downward by 1~km, 
while displacing the exit profile upward by the same amount 
(because the two errors and anticorrelated, see \citealt{hin17}) 
changes the relative amplitude of the flash by a mere $10^{-3}$, 
well below the noise level of our observations (Fig.~\ref{fig_fit_rex_boo_dun}).
In other words, our central flash observations cannot pin down
the absolute vertical scales of the profiles to within the $\pm 3.6$~km REX uncertainty.

The fits are displayed in Fig.~\ref{fig_fit_rex_boo_dun}.
Their qualities are estimated through the $\chi ^2$ value.
Depending on the fits, there are $M=1$ to 3 free parameters
(the pressure at a prescribed level, off-track displacement of Pluto
with respect to its ephemeris and Pluto's contribution $\phi_P$ to the flux).
In all the fits, there are $N=217$ data points adjusted.
Note that the value of $h$ in Case (4) has been fixed to 1.35~km,
i.e. is not an adjustable parameter. This is discussed further in the points below:
\begin{enumerate}
\item
The nominal temperature profile $T(r)$ of \cite{sic16} with surface pressure $p_{\rm surf}=12.7$~$\mu$bar
provides a satisfactory fit with $\chi^2=198$ ($\chi^2_{\rm dof}= \chi^2/(N-M)= 0.924$ per degree of freedom).
In this case, the Bootes-3 and Dunedin stations passed 46~km north and 45~km south of the shadow center, respectively.
\item
The nominal REX profiles result in flashes that are too high compared to the observations,
as noted by a visual inspection of the figure
(and from $\chi^2=326$, $\chi^2_{\rm dof}=1.52$).
This can be fixed by introducing haze absorption. 
A typical factor of 0.7 must be applied to the Bootes-3 synthetic flash in order to match the data,
while a typical factor of 0.76 must be applied to the Dunedin synthetic flash. 
This corresponds to typical tangential optical depths (along the line of sight) in the range $\tau_T = 0.27-0.35$,
for rays that went at about 8~km above the REX 1187.4~km radius.
Changing Pluto's off-track offset does not help in this case, as one synthetic flash increases
while the other decreases. 
This could be accommodated by adjusting accordingly the optical depths $\tau_T$,
but this introduces too many adjustable parameters to be relevant.
\item
A satisfactory best fit is obtained ($\chi^2=214$, $\chi^2_{\rm dof}=0.999$)
by reducing uniformly the REX density profiles by a factor of 0.805 and by
moving Pluto's shadow center cross-track by 17~km north with respect to Case (1), 
the Bootes-3 and Dunedin stations passing 29~km north and 62~km south of the shadow center, respectively.
This displacement corresponds to a formal disagreement at 3-$\sigma$ level 
for Pluto's center position between Case (1) and (3),
when accounting for the noise present in the central flashes (Fig.~\ref{fig_fit_rex_boo_dun}).
Thus, such difference  remains marginally significant. 
Note also that a satisfactory fit to the Bootes-3 flash is obtained, while the Dunedin synthetic flash remains a bit too high. 
As commented in the concluding Section, however, a reduction of the density profile by a factor of 0.805
is implausible considering the error bars of the REX profiles.
\item
Using again the nominal REX profiles of Case (2), but imposing a topographic feature of height $h=1.35$~km
on top of the REX exit radius of 1192.4~km, a satisfactory fit to the Bootes-3 flash is obtained
($\chi^2=205$, $\chi^2_{\rm dof}=0.959$),
in fact the best of all fits for that station. 
Meanwhile, the Dunedin synthetic flash remains a bit too high compared to observations.
In this model, Pluto's shadow center has been moved cross-track by 19.5~km 
north with respect to the first model, 
so that the Bootes-3 and Dunedin stations passed 26.5~km north and 64.5~km south of the shadow center, respectively.
Again the discrepancy relative to the Pluto's center solution of Case (1) is at 3-$\sigma$ level, and thus marginally significant.
The particular choice of $h=1.35$~km stems from the fact that lower values would increase even more
the Dunedin flash, while higher values would decrease too much the Bootes-3 flash.
We have not explored further other values of $h$ by tweaking the density profiles. 
So, this is again an exercise to show that reasonably high topographic features may explain the observed flash.
\end{enumerate}

\section{Concluding remarks}
\label{sec_conclusion}

\subsection{Pluto's global atmospheric evolution}

Fig.~\ref{fig_pressure_time} summarizes our results 
concerning the evolution of Pluto's atmospheric pressure with time.
It shows that the observed trend can be explained by adjusting Pluto's physical parameters
in a rather restrictive way.

As noted in Section~\ref{sec_pres_evolution},  
this evolution is consistent with the continuous increase of pressure observed since 1988 
(a factor of almost three between 1988 and 2016). 
It results from the heating of the nitrogen ice in Sputnik Planitia and
in the northern mid-latitudes, when the areas are exposed to the Sun (just after the northern spring equinox in 1989) and
when Pluto is near the Sun \citep{ber16}.
The model also predicts that atmospheric pressure is expected to reach its peak and drop in the next few years, due to 
  
  (1) the orbitally-driven decline of insolation over Sputnik Planitia and the northern mid-latitude deposits, and 
  
  (2) the fact that nitrogen condenses more intensely in the colder southern part of Sputnik Planitia, 
  thus precipitating and hastening the pressure drop. 
  
In that context, it is important to continue the monitoring of Pluto's atmosphere using ground-based stellar occultations.
Unfortunately, as Pluto moves away from the Galactic plane, such occultations will become rarer and rarer.

\subsection{Pluto's lower atmosphere}

The models presented in the Section~\ref{sec_lower_atmo} and illustrated in Fig.~\ref{fig_fit_rex_boo_dun}
are not unique and not mutually exclusive. 
For instance, one can have at the same time a topographic feature blocking the stellar rays,
together with some haze absorption. 
Also, hazes, if present, will not be uniformly distributed along the limb.
Similarly, topographic features will probably not be uniformly distributed along the limb,
but rather, have a patchy structure that complicates our analysis.
In spite of their limitations, 
the simple scenarios presented above teach us a few lessons: 

(1) Although satisfactory in terms of flash fitting, the nominal temperature profile of \cite{sic16}
seems to be ruled out below the planetocentric radius $\sim 1215$~km, 
since it is clearly at variance with the REX profiles (Fig.~\ref{fig_T_r_rex_compa_bs}),
while probing essentially the same zones on Pluto's surface (Fig.~\ref{fig_lon_lat_boot_dun}). 
As discussed in Section 4.2 however, diurnal changes occurring over Sputnik Planitia might explain this discrepancy, 
with a cooler (sunset) REX temperature profile and a warmer (sunrise) profile more in line with the DO15 solution.
However, current GCM models predict that these diurnal changes should occur below the 5-km altitude level,
and not as high as the 25~km observed here. 
This issue remains an open question that would be worth investigating in future GCM models.

(2) The REX profiles taken at face value cannot explain the central flashes observed at Bootes-3 and Dunedin,
unless hazes are present around the $\sim 8$~km altitude level, 
with optical depths along the line of sight in the range $\tau =$ 0.27-0.35.
This is higher but consistent with the reported value of $\tau \sim 0.24$ 
derived from NH image analysis \citep{gla16,che17}.
In fact, the two values are obtained by using quite different methods.
\cite{che17} assume tholin-like optical constant, which is not guaranteed. 
Moreover, their 0.24 value is the scattering optical depth,
while we measure the aerosol extinction (absorption plus scattering).
Chromatic effects might also be considered to explain those discrepancies,
as the Bootes-3, Dunedin and the NH instruments have different spectral responses.
Our data are too fragmentary, though, to permit such a discussion.

(3) An alternative solution is to reduce uniformly the REX density profiles by a factor 0.805.
However, this would induce a large disagreement (8-$\sigma$ level) on the REX density profile at 7~km altitude,
and thus appears to be an unrealistic scenario.
%
%
Moreover, the underdense versions of the REX profiles would then disagree formally 
(i.e. beyond the internal error bars of the DO15 light curve fitting model)
when extrapolated to the overlying half-light level around $r=1300$~km.
A remedy would be to patch up ground-based-derived profiles 
with the underdense REX profiles, and re-run global fits.
This remains out of the scope of the present analysis.

(4) The topographic feature hypothesis remains an attractive alternative, 
as it requires modest elevation (a bit more than 1~km) above the REX exit region,
that is known to be higher than the entry region, Sputnik Planitia.
A more detailed examination of Pluto's elevation maps, 
confronted with the stellar paths shown in Fig.~\ref{fig_lon_lat_boot_dun},
should be undertaken to confirm or reject that hypothesis.
This said, such $\pm$~1 km topographic variations are actually observed 
all over Pluto's surface \citep{sch18b}.

As a final comment, we recall that the flashes have been generated 
by assuming a spherical atmosphere near Pluto's surface. 
There is no sign of distortion of the Bootes-3 and Dunedin flashes that suggests a departure from sphericity. 
It would be useful, however to assess such departures, 
or at least establish an upper limit for them in future works.

\begin{acknowledgements}
This article is dedicated to the memory of H.-J. Bode, J.~G. Greenhill and O. Farag\'o 
for their long-standing support and participation to occultation campaigns.
The work leading to these results has received funding from the 
European Research Council under the European Community's H2020
2014-2020 ERC Grant Agreement n$^\circ$ 669416 ``Lucky Star".
EM thanks support from Concytec-Fondecyt-PE and GA, FC-UNI for providing support during the 2012 July 18 occultation.
BS thanks 
S. Para for partly supporting this research though a donation, 
J.~P. Beaulieu for helping us accessing to the Hobart Observatory facilities
and B. Warner, B. L. Gary, C. Erickson, H. Reitsema, L. Albert, P. J. Merritt, T. Hall, W. J. Romanishin, Y. J. Choi  
for providing data during the 2007 March 18 occultation.
MA thanks CNPq  (Grants 427700/2018-3, 310683/2017-3 and 473002/2013-2) and FAPERJ (Grant E-26/111.488/2013).
JLO thanks support from grant AYA2017-89637-R.
PSS acknowledges financial support from the European Union's Horizon 2020 Research and Innovation
Programme, under Grant Agreement no 687378, as part of the project ``Small Bodies Near and Far" (SBNAF).
JLO, RD, PSS and NM acknowledge financial support from the State Agency for Research of the Spanish MCIU 
through the ``Center of Excellence Severo Ochoa" award for the Instituto de Astrof\'{\i}sica de Andaluc\'{\i}a (SEV-2017-0709).
FBR acknowledges CNPq support process 309578/2017-5.
GBR thanks support from the grant CAPES-FAPERJ/PAPDRJ (E26/203.173/2016).
JIBC acknowledges CNPq grant 308150/2016-3.
RVM thanks the grants: CNPq-304544/2017-5, 401903/2016-8, and 
Faperj: PAPDRJ-45/2013 and E-26/203.026/2015.
BM thanks the CAPES/Cofecub-394/2016-05 grant and CAPES/Brazil - Finance Code 001.
BM and ARGJ were financed in part by the Coordena\c{c}\~ao de Aperfei\c{c}oamento de 
Pessoal de N\'{\i}vel Superior - Brasil (CAPES) - Finance Code 001.
TRAPPIST-North is a project funded by the University of Li\`ege, 
in collaboration with Cadi Ayyad University of Marrakech (Morocco).
TRAPPIST-South is a project funded by the Belgian Fonds (National) de la Recherche Scientifique
(F.R.S.-FNRS) under grant FRFC 2.5.594.09.F, with the participation of the Swiss National Science
Foundation (FNS/SNSF).
VSD, SPL, TRM and ULTRACAM are all supported by the STFC.
KG acknowledges help from the team of Archenhold-Observatory, Berlin, 
and AR thanks G. Rom\'an (Granada) for help during the observation of the 2016 July 19 occultation.
AJCT acknowledges support from the Spanish Ministry Project AYA2015-71718-R (including EU funds).
We thank Caisey Harlingten for the repeated use of his 50 cm telescopes in San Pedro de Atacama, Chile.
We thank the Italian Telescopio Nazionale Galileo (TNG), 
operated on the island of La Palma by the Fundaci\'on Galileo Galilei of the INAF (Istituto Nazionale di Astrofisica) 
at the Spanish Observatorio del Roque de los Muchachos of the Instituto de Astrof\'{\i}sica de Canarias.
LM acknowledges support from the Italian Minister of Instruction,
University and Research (MIUR) through FFABR 2017 fund and
support from the University of Rome Tor Vergata through ``Mission: Sustainability 2016'' fund.
The Astronomical Observatory of the Autonomous Region of the Aosta Valley (OAVdA) is managed by 
the Fondazione Cl\'ement Fillietroz-ONLUS, which is supported by the Regional Government of the Aosta Valley, 
the Town Municipality of Nus and the ``Unit\'e des Communes vald\^otaines Mont-\'Emilius". 
The research was partially funded by a 2016 ``Research and Education" grant from Fondazione CRT.
We thank D.P. Hinson for his constructive and detailed comments that helped to improve this article.
\end{acknowledgements}




\section{Circumstances of Observations}


\longtab{
\begin{longtable}{lllll}
\caption{
\label{tab_sites}
Circumstances of Observations
} \\
\multicolumn{5}{c}{\textbf{DATE}} \\
\textbf{Site} & \textbf{Coordinates}   & \textbf{Telescope} & \textbf{Exp. Time/Cycle (s)} &\textbf{Observers} \\
	           &    \textbf{altitude (m)} &   \textbf{Instrument/filter}    &     &  \\
\hline
\hline
\endfirsthead
\multicolumn{5}{c}%
{\tablename\ \thetable\ -- \textit{Continued from previous page}} \\
\multicolumn{5}{c}{\textbf{DATE}} \\
\textbf{Site} & \textbf{Coordinates}   & \textbf{Telescope} & \textbf{Exp. Time/Cycle (s)} &\textbf{Observers} \\
	           &    \textbf{altitude (m)} &   \textbf{Instrument/filter}    &     &  \\
\hline
\hline
\endhead
\hline \multicolumn{5}{r}{\textit{Continued on next page}} \\
\endfoot
\hline
\endlastfoot
\multicolumn{5}{c}{\textbf{2002 August 21}} \\
\hline
CFHT           &  19 49  30.88     N     &  3.6m                                     &   1/1.583      & C. Veillet  \\
			Hawaii          &  155 28 07.52  W      &  I ($0.83 \pm 0.1$~$\mu$m)  &                    &                  \\
			                     &  4200                        &                                               &                     &                 \\
\hline
\multicolumn{5}{c}{\textbf{2007 June 14}} \\
\hline
Pico dos Dias   &  22 32 7.80     S       &  1.6m               &    0.4/0.4        & F. Braga-Ribas,    \\
Brazil                &  45 34 57.70  W    &   CCD/clear        &                       & D. Silva Neto          \\
                         &    1864                  &                                  &                        &            \\
			                        
			Hakos           &  23 14 50.4 S         &   IAS 0.5m          &   1.373/1.373        & M. Kretlow  \\
			Namibia        & 16  21 41.5  E       &   TC245 IOC/clear &          &             \\
			                       &    1825.                &                         &                        &            \\ 

			Paranal         & 24 37 39.44 S      &   UT1 8.2m                &  0.1/0.1     &   V. Dhillon,       \\
			Chile               & 70 24 18.27  W    &  Ultracam/u',g,'i'     &            &   S. Littlefair,         \\
			                      & 2635                     &                             &                     &  A. Doressoundiram \\
			Paranal          & 29 15 16.59 S      &   VLT Yepun 8.2m    &    1/1      & B. Sicardy   \\  
			Chile               & 70  44 21.82  W  & NACO/Ks  &                            &         \\
			                      & 2315.                  &                   &                     &                          \\
\hline
\multicolumn{5}{c}{\textbf{2008 June 22}} \\
\hline
			Bankstown     &  33 55 56  S     &  0.275m                            &  1.28/1.28  & T. Dobosz \\
	                Australia& 151 01 45 E    & video/clear         &      &  \\
	                & 24.9                 &     &     & \\

			Blue Mountains  &   33 39 51.9 S   & 0.25m                              &  1.28/1.28   & D. Gault \\ 	                   
                          Australia  	  & 150 38 27.9 E  &  video/clear       &             & \\           
                            	  & 286                    &   &         & \\

			Reedy Creek      &   28 06 29.9 S  & 0.25m             & 6.30/8.82    & J. Broughton  \\  
                            Australia	  & 153 23 52.0 E  & CCD/clear         &   & \\           
                            	  &  65                     &   &   & \\

	                    Glenlee                &   23 16 09.6 S  & 0.30m                                & 0.12/012  & S. Kerr \\ 	                   
                            Australia	  & 150 30 00.8 E  &  video/clear        &                         & \\           
                            	  & 50                      &     &                           & \\

			Perth  &  31 47 21.5 S    & 0.25m                              &     & G. Bolt\\ 	                   
                         Australia   	     &115 45 31.3 E   &  CCD/clear        &  2.0  & \\           
                                &  45                      &   &  6.0   & \\                             
\hline
\multicolumn{5}{c}{\textbf{2008 June 24}} \\
\hline
CFHT           &  19 49  30.88     N       &  3.6m           &   0.065/0.065  &  L. Albert  \\
			Hawaii                  &  155 28 07.52  W    &   Wircam/K          &             \\
			                        &    4200                &                         &                        &            \\
\hline
\multicolumn{5}{c}{\textbf{2010 February 14}} \\
\hline
Pic du Midi     &  42 56 12.0  N     &  T1m                           &  0.32/0.32  & J. Lecacheux \\
	                France            &  00 08 31.9 E    & CCD/clear          &      &  \\
	                                       & 2862                &     &     & \\

			Lu                 &   46 37 26.3 N    & 0.35m          &  0.35/0.50   & C. Olkin, \\ 	                   
                         Switzerland  & 10 22 00.3 E      &  video/clear    &                     & L. Wasserman\\           
                            	            & 1933                  &                        &                    & \\

			Sisteron      &  44 05 18.20 N  & 0.3m                     & 0.64/0.64   & F. Vachier \\  
                          France	  &  05 56 16.3  E   & Watec 120/clear     &   & \\           
                            	          &  634                     &   &   & \\     
\hline
\multicolumn{5}{c}{\textbf{2010 June 04}} \\
\hline
Mt John          &  43 59 13.6 S    &  1m                  &  0.32/0.32   & B. Loader,\\ 	                   
New Zealand &170 27 50.2 E    &  CCD/clear        &    & A. Gilmore, P. Kilmartin \\           
                      &  1020                 &   &      &  \\  

			Hobart &  42 50 49.83 S      &  1m                            &  1/1    &  J. G. Greenhill, \\ 	                   
                         Australia &147 25 55.32 E   &  Raptor/I                      &            &  S. Mathers \\           
                                       &  38                   &                                      &           &  \\    
	                                      
	               	Blenheim &  41 29 36.3 S        & Bootes-3 0.6m   &  0.50/1.75   & W. H. Allen \\ 	                   
                         New Zealand &173 50 20.7 E  &  CCD/r'                   &                     & \\           
                                       &  37.5                     &   &      & \\     	                        
			
			Blenheim &  41 29 36.3 S         &  0.4m               &  2.5/6    & W. H. Allen \\ 	                   
                         New Zealand &173 50 20.7 E   &  CCD/clear            &             & \\           
                                       &  37.5                        &   &      & \\    
	                 
	                 Oxford           &  43 18 36.78  S     &  0.3m                             &  0.64/0.64 & S. Parker \\
	                 New Zealand & 172 13 07.8 E       &   Video/clear                    &      &  \\
	                                       & 221                &     &     & \\
\hline
\multicolumn{5}{c}{{\textbf{2011 June 04}}} \\
\hline
Santa Martina       &  33 16 09.0     S       &  0.4m                &   2/2        & R. Leiva     \\
Chile                    &  45 34 57.70  W        &  EMCCD/clear &          &           \\
                            &    1450                      &                         &                        &            \\

La Silla            & 29 15 16.59 S     & TRAPPIST S 0.6m  & 3/4.4        & E. Jehin                 \\
Chile                & 70  44 21.82  W  & CCD/clear                &                 &                                \\
                        & 2315                   &                                 &                   &                                 \\

San Pedro de    &   22 57 12.3 S       & Caisey 0.5m      & 2/2.87                     & A. Maury \\
Atacama, Chile  &   68 10 47.6 W       &  CCD/clear                                &    &                   \\
                          &   2397                   &                                         &     &             \\

Pico dos Dias   &  22 32 7.80     S       &  1.6m                      &    0.1/0.1        & M. Assafin     \\
Brazil                &  45 34 57.70  W    &   CCD/clear        &                       &           \\
                         &   1864                  &                                  &                        &            \\                      
\hline
\multicolumn{5}{c}{\textbf{2012 July 18}} \\
\hline
Santa Martina       &  33 16 09.0     S       &  0.4m           &   1/1        & R. Leiva     \\
Chile                    &  45 34 57.70  W    &   CCD/clear &          &           \\
                            &    1450               &                         &                        &            \\
			                        
Cerro Burek           &  31 47 12.4 S     &   ASH 0.45m          &  13/15.7        & N. Morales  \\
Argentina                       & 69  18 24.5  E    &   CCD/clear &          &             \\
			                        &    2591                &                         &                        &            \\ 

Paranal         & 24 37 31.0 S      &   VLT Yepun 8.2m             & 0.2/0.2     & J. Girard       \\
Chile               & 70 24 08.0  W    &  NACO/H   &                    &             \\
			                      & 2635                     &                      &                   &   \\

San Pedro de       & 22 57 12.3 S         &   ASH2 0.4m    &    13/15.44   & N. Morales   \\  
Atacama, Chile   & 68  10 47.6  W  & CCD/clear  &                      &       \\
			                            & 2397                  &                   &                     &                          \\
			 
Huancayo         & 12 02 32.2 S      &   0.20m    &    10.24/10.24   & E. Meza   \\  
Peru               & 75 19 14.7  W  & CCD/clear  &    5.12/5.12                  &       \\
			                        & 3344                 &                   &                     &                          \\
\hline
\multicolumn{5}{c}{\textbf{2013 May 04}} \\
\hline
Pico dos Dias & 22 32 07.8 S       &  B\&C 0.6m          &        4.5/6        &  M. Assafin,          \\
Brazil               & 45  34 57.7  W    &              CCD/I                 &           &   A. R. Gomes-J\'unior                   \\
                         & 1,811                      &                             &                                  &                                                                    \\

Cerro Burek    &  31 47 14.5 S   & ASH 0.45 m               &  6/8 & J.L. Ortiz \\
Argentina        &  69 18 25.9 W   &                 CCD/clear                 &        &  \\
                         & 2591              &                                  &        &  \\

Cerro Tololo   & 30 10 03.36 S      & PROMPT 0.4m   & 5/8                               &   J. Pollock               \\
Chile              & 70 48 19.01  W    & P1, P3, P4, P5    &  P3 offset 2 sec          &            \\
                         & 2207                     &       CCD/clear         &  P4 offset 4 sec          &             \\
                         &                                &                               &  P5 offset 6 sec          &                                    \\

La Silla           &  29 15 21.276 S   & Danish 1.54m           & Lucky Imager     & L. Mancini  \\
Chile              & 70 44 20.184  W  &         Lucky Imager/Z  ($>$650nm    & 0.1/0.1               &                    \\
                        &  2336                     &       CCD/iXon response)                  &                          &                     \\

La Silla           & 29 15 16.59 S    & TRAPPIST S 0.6m  & 4.5/6           & E. Jehin                 \\  
Chile               & 70  44 21.82  W  &           CCD/clear                      &   &                                \\
                        & 2315                     &                                 &                 &                                 \\

Cerro Paranal & 24 37 31.0   S      & VLT Yepun 8.2m  &   0.2/0.2                        &   G. Hau          \\
Chile              & 70 24 08.0    W     &   NACO/H               &                        &    \\
                           & 2635.43              &                                   &                                &       \\

San Pedro de      &   22 57 12.3 S       & Caisey 0.5m f/8    & 3/4.58                     & A. Maury \\
Atacama, Chile    &   68 10 47.6 W       & CCD/V                                 &    &                   \\
                                                     &   2397                    &                                         &     &             \\
                    
San Pedro de   &22 57 12.3 S          & Caisey 0.5m f/6.8       &4/4.905             & L. Nagy \\
Atacama, Chile &      68 10 47.6 W     &  CCD/B                           &  &                \\
                        &                                 &                                         &           &                \\
\hline
\multicolumn{5}{c}{\textbf{2015 June 29}} \\
\hline
Lauder  & 45 02 17.39 S         &  Bootes-3/YA 0.60m  & 0.05633/0.05728 &   M. Jel\'{\i}nek  \\
New Zealand   & 169 41 00.88 W      &   EMCCD/clear    &   &  \it central flash detected                   \\
 	                &   382 	&	         &	&  \\

Dunedin                &    45  54  31 S    &   0.35m   & 5.12/5.12         &   A. Pennell,  S. Todd,      \\
New Zealand         & 170  28  46  E    &   CCD/clear                        &            &  M. Harnisch, R. Jansen        \\
                                            & 136           &                                         &          &  \it central flash detected \\  

Darfield                 &    43  28  52.90 S    &   0.25m          & 0.32/0.32         &    B. Loader       \\
New Zealand         & 172  06  24.40 E    &   CCD/clear    &                        &    \it central flash detected  \\
                                            & 210                    &                                         &          &  \\
		                                    
Blenheim  1            &   41  32  08.60  S    &   0.28m   & 0.64/0.64         &    G. McKay       \\
New Zealand         & 173  57  25.10 E    &   CCD/clear               &          &            \\
                                            & 18                    &                                         &          &   \\

Blenheim   2            &    41  29 36.27 S    &  0.4m     &       0.32/0.32       &    W. H.  Allen   \\
New Zealand         &  173  50 20.72 E    &   CCD/clear               &          &            \\
                                            & 38                    &                                         &          &   \\   

Martinborough       &    41  14  17.04 S    &  0.25m    &       0.16/0.16       &    P. B.  Graham   \\
New Zealand         &  175  29  01.18 E    &   CCD/B                  &          &            \\
                                            & 73                   &                                         &          &  \\                                     

Greenhill Obs.  &   42  25 51.80 S    &   1.27m  &       0.1/0.1       & A. A. Cole,      \\ 
Australia                  & 147  17  15.80 E   &   EMCCD/B                &                         &  A. B. Giles,       \\
                                            & 641                     &                                         &          &  K. M. Hill \\

Melbourne          &    37  50  38.50 S    &  0.20m    &       0.32/0.32       &    J.  Milner   \\
Australia             &  145  14 24.40 E    &   CCD/clear               &          &            \\
                            & 110                    &                                         &          &   \\ 
\hline
\multicolumn{5}{c}{\textbf{2016 July 19}} \\
\hline
Pic du Midi  & 42 56 12.0 N         &  1m    & 0.3/0.3 &   F. Colas,  \\
France        & 00 08 31.9 E      &   EMCCD/clear       &   &  E. Meza                   \\
 	                           &   2862 	&	         &	& \\

Valle d'Aosta  &  45 47 22.00 N       &  0.81m   & 1/1     &   B. Sicardy,  \\
Italy          &       7 28 42.00 E      &   EMCCD/clear  &   &  A. Carbognani                   \\
 	                &   1674 	&	         &	& \\ 

La Palma  &  28 45 14.4 N       & TNG 3.58m        & 1/5     &  L. di Fabrizio,  A. Magazz\'u,   \\
Spain          & 17 53 20.6 E      &   EMCCD/clear   &           &  V. Lorenzi,  E. Molinari     \\
 	                &   2387.2 	&	         &	& \\                                              

Saint V\'eran  &  44 41 49.88 N    &  0.5m                    & 0.3/0.3          &   J.-E. Communal,    \\
France           &   06 54 25.90 E   &  EMCCD/clear       &                &   S. de Visscher,  F. Jabet,        \\
 	             &   2936                  &  0.62m                    & 0.2/0.2   &   J. S\'erot   \\   
                     &                              & near IR camera/    &                &    \\
 	             &                           &  RG 850 long pass   &                &    \\   
                                          
Calern  &  43 45 13.50 N      &  C2PU T1m & 0.3/0.3   &  D. Vernet,  J.-P. Rivet,            \\
France   & 06 55 21.80 E     &   EMCCD/clear  &              & Ph. Bendjoya, M. Devog\`ele   \\
 	      &   1264 &  &  & \\          

Mitzpe Ramon  &   30 35 44.40 N    &  Jay Baum Rich   & 1/2.5   &  S. Kaspi, D. Polishook,  \\
Israel                &  34 45 45.00 E     &  Telescope 0.7m                &           & N. Brosh, I. Manulis  \\
 	                &  862                      &  CCD/clear                        &           & \\  	
Trebur              & 49 55 31.6 N         & T1T 1.2m               & 0.3/0.3 &	 J. Ohlert \\
Germany         & 08 24 41.1 E          & CMOS/clear &          & \\
                        & 90 & & &\\ 

Athens  & 37 58 06.8 N & 0.4m          & 2/4.5   & K. Gazeas, \\ 
Greece & 23 47 00.1 E  & CCD/clear &                &  L.Tzouganatos \\   
 	            &  250                 &                          &                & \\ 
Ellinogermaniki   & 37 59 51.7 N  & 0.4m          & 7/11 & V. Tsamis,  \\                                                                                  
Agogi, Pallini &  23 58 36.2 E  & CCD/clear &                     & K.Tigani \\
Greece &  169                 &                          &               & \\  	
                  
\hline\hline
\multicolumn{5}{c}{\textbf{Data sets not included in this work}} \\
\hline\hline
\multicolumn{5}{c}{\textbf{2002 July 20}} \\
\hline
Arica           &  18 26 53.8 S    &  0.3m             &   2/2     & F. Colas  \\
Chile          &  69 45 51.5 W   &  CCD/clear      &             &                 \\
                  &  2500                &                         &             &                 \\
\hline
\multicolumn{5}{c}{{\textbf{2006 June 12}}} \\
\hline

Stockport              &  34 19 55.31 S           &  0.50m        &  1.5/2       & B. Lade    \\
Australia                &   138 43 45.38 E        &  CCD/clear  &                 &                  \\
			     &   24                             &                   &                  &                 \\
			                     
Blue Montains    &  33 39 51.9 S     &  0.25m           &  1/2            & D. Gault \\
Australia            & 150 38 27.9 E    & CCD/clear    &                     &                  \\
                         &  286                    &                      &                     &                 \\
			                     
Hobart     &   42 50 49.83 S     &  0.4m              &   1.6/1.6      & W. Beisker, \\
Australia  &    147 25 55.32 E    &                         &                 &  A. Doressoundiram,     \\
                &    38                        &                         &                  &  S. W. Dieters, J. G. Greenhill  \\
\hline
\multicolumn{5}{c}{{\textbf{2007 March 18}}} \\
\hline
Catalina Mts.       & 32 25 00 N     &  Kuiper 1.53m     &  0.68/0.68    & T. Widemann \\
USA                     & 110 43 57 W   &  CCD/clear  &                    &                  \\
                            &  2790              &                    &                     &                 \\		                     

Palmer  Divide      & 39 05 05 N       &  0.35m                               &  16.9/16.9     &  B. Warner \\
USA			    & 104 45 04  W   &  CCD/clear &                    &                  \\
                             &  2302                     &                                               &                     &                 \\
			                     
Calvin Rehoboth         &  35 31 32 N     &   0.4m                       &   8.5/8.5     &   L. A. Molnar \\
USA			           & 108 39 23  W  &  CCD/I  &                    &                  \\
			           &  2024                      &                                               &                     &                 \\

Cloudbait                    &  38 47 10 N         & 0.305m              &  29/29      & C. Peterson \\
USA			          &  105 29 01 W       & CCD/clear       &                    &                  \\
			          &  2767                    &                        &                     &                 \\	

Hereford                    & 31 27 08 N         & 0.36m                 &  3/5.1      & B. Gary  \\
USA			         & 110 14 16 W      & CCD/clear           &                    &                  \\
			         &  1420                  &                            &                     &                 \\

Oklahoma                 &  35 12 09 N      & 0.4m                                &  4/6.2     & W. Romanishin\\
USA			        &  97 26 39 W     & CCD/R+I    &                    &                  \\
                                 &  382                 &                                           &                     &                 \\

Mt  Lemmon             &  32 26 32 N        &  Kasi 1m       &   17.6/17.6      & Y.-J. Choi  \\
USA			         & 110 47 19 W      & CCD/I            &                        &           \\
			         &  2776                        &                                           &                     &           \\      
\hline
\multicolumn{5}{c}{{\textbf{2007 June 09}}} \\
\hline			                     
		
Cerro Pach\'on    & 30 14 16.80 S      & SOAR 4.1m        &  0.66/0.66   & W. Beisker \\
Chile                   &  70 44 1.35 W      & CCD/dual B \& R  &                   &     \\
                          & 2715                    &   &   &      \\			                     

\hline
\multicolumn{5}{c}{{\textbf{2008 August 25}}} \\
\hline
Lick                & 37 20 24.6         &  Shane 3.0m    &  0.8/0.8    & F. Marchis \\
USA               & 121 38 43.8       & IR mosaic/K            &                    &                  \\
                      & 1281                   &                                               &                     &                 \\
			                     
Grands Rapids  & 42 55 50 N     &   0.4m    &  10/13.3      &  L. A. Molnar \\
USA                   & 85 35 18 W     &  CCD/I   &                    &                  \\
                          &  253                 &               &                     &                 \\
\hline
\multicolumn{5}{c}{{\textbf{2010 May 19}}} \\
\hline
Paranal  & 24 37 36.64 S    & VLT Melipal 8.2m      &  0.5/0.5      & B. Sicardy \\
Chile      & 70 24 16.32 W    & ISAAC/Ks &   &      \\
              & 2635                    &   &   &      \\

La Silla     &  29 15 32.1 S    &  NTT 3.58m    &  0.5/0.5      &V. D. Ivanov \\
Chile        &  70 44 0.15 W   & SOFI/Ks            &   &      \\
                &  2375                &                           &   &      \\
			 
Cerro Pach\'on    & 30 14 16.80 S      & SOAR 4.1m       &  2.5/3.5      & M. Assafin \\
Chile                  & 70 44 1.35 W        & CCD/clear &  &      \\
                          &  2715                    &   &   &      \\
\hline
\multicolumn{5}{c}{{\textbf{2011 June 23}}} \\
\hline
San Pedro M\'artir &  31 02 39  N   & 2.1m             &       1/1.52             & R. Howell \\
Mexico                 & 115 27 49 W   & IR mosaic/K  &                               &                    \\
                            & 2800 m           &                        &                              &                    \\

San Pedro M\'artir &  31 02 43.1  N   & 0.84m                 &   0.35/0.35             &  R. French \\
Mexico                 & 115 27 57.7 W    & CCD/clear          &    &                     \\
                             & 2811 m               &                           &         &                    \\

Hale A'a  BB        &  19 09 29.6  N     & 0.6m                           &  1/1                 &  E. Young   \\
                               & 155 45 19.1 W  &  CCD/clear                &                         &             \\
                               & 1509 m             &                                    &                    &                              \\

Hale A'a   CE        &  19 09 29.6  N    & 0.4m                 &    1/1                   &  C. Erickson  \\
                               & 155 45 19.1 W  &  CCD/clear        &                           &                        \\
                               & 1509 m             &                           &                             &                           \\

Haleakala              &  20 42 27.0  N    &  FTN 2m      &     0.093/0.09974       &  F. Bianco       \\
                               & 156 15 21.0 W  &  CCD/I          &                                   &        \\
                               & 3055 m             &                      &                                    &                         \\

Kekaha                &  21 58 15.15  N       & 0.4m          &  0.3/0.3                 &  T. Widemann,            \\
                            & 159 43 21.558 W     &  CCD/clear      &                        &  M. Buie, T. Hall          \\
                            & 20 m                        &    &         &                            \\

KEASA                &  21 59 05.7  N        & 0.35m                         &   0.333/0.333               & J. Merrit    \\
                               & 159 45 09.8 W    & CCD/clear                  &                                     &          \\
                               & 10 m                    &                                    &                                    &            \\
                               
Maui                      &  20 54 43.2  N        & 0.35m                         &   1/1               & H.-J. Bode    \\
                               & 156 41 28.9 W     & CCD/clear                  &     &             \\
                               & 47 m                    & (partly cloudy)              &                           &             \\

Majuro                  &   07  04   06.6  N    & 0.4m                   &     0.8/0.8          & C. Olkin,    \\
                               & 171 17 39.8 W      &  CCD/I                &                         & H. Reitsema      \\
                               & 8 m                       &                            &                          &                           \\
\hline
\multicolumn{5}{c}{{\textbf{2012 June 14}}} \\
\hline
Marrakech     &  31 35 16.2  N   &  0.6m                  &       0.5/0.5             & S. Renner, Z. Benkhaldoun,  \\
Morocco         & 08 00 46.9 W   &  EMCCD/clear     &                             & M. Ait Moulay Larbi,       \\
                      & 494 m            &                               &                              & A. Daassou, Y. El Azhari                   \\                  

Sierra Nevada  &  37 03 51  N                &  1.52m                       &  1.5/2           & J.~L. Ortiz                         \\
Obs., Spain      & 03 23 49  W                 &  CCD/clear               &                   &          \\
                        & 2925                            &                                  &                     &       \\

\hline
\multicolumn{5}{c}{{\textbf{2016 July 14}}} \\
\hline
Oukame\"iden     &  31 12 23.2  N   &  TRAPPIST N 0.6m      &       2/3            & E.  Jehin  \\ 
Morocco             & 07 51 59.3 W   &  CCD/clear                     &                        &                    \\
                           & 2720 m            &                                        &                        &                    \\                  
                   
Sierra Nevada   &  37 03 51  N                &  0.9m                       &  2/3.5     & J.~L. Ortiz                         \\
Obs., Spain       & 03 23 49  W                   &  CCD/clear               &                   &          \\
                        & 2925                            &                                  &                     &       \\

Granada         &  36 59 33.2 N               &  Dobson 0.6m        &  3.5/3.5          & S. Alonso, D. B\'erard,   \\
Spain               &  03 43 19.9 W             &  CCD/clear            &                       & A. Rom\'an         \\
                        & 1130                             &                              &                       &        \\

\hline                                     
\end{longtable}
}

\begin{appendix}

\section{Reconstructed geometries of the occultations}

\begin{figure*}
\centerline{%
\includegraphics[totalheight=0048mm,trim=0 0 0 0,angle=0]{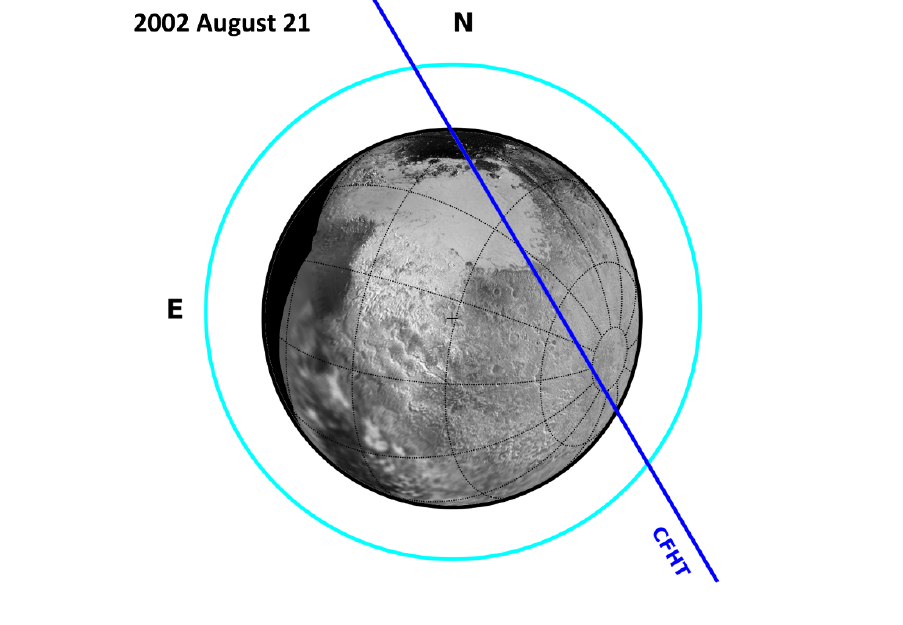}
\includegraphics[totalheight=0048mm,trim=0 0 0 0,angle=0]{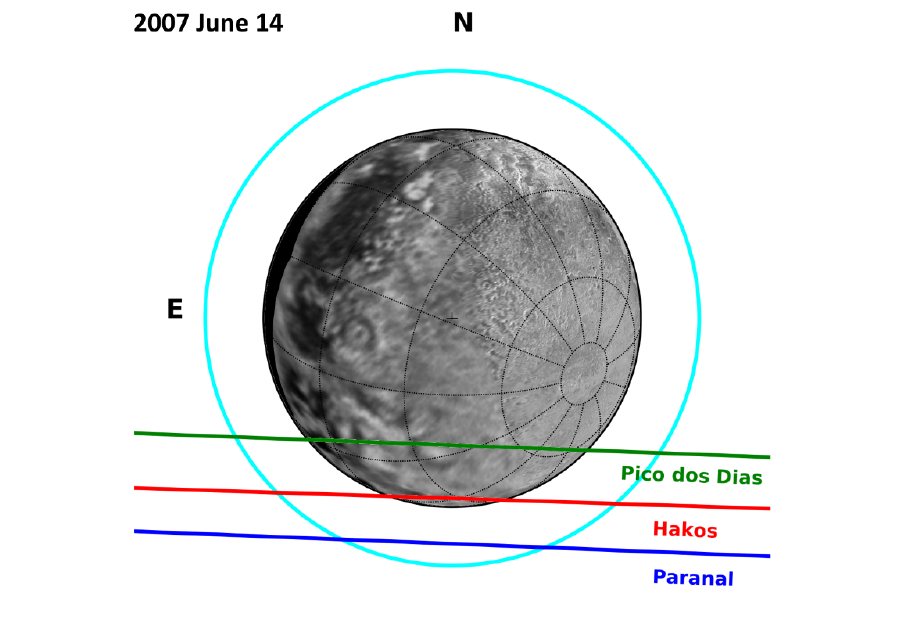}
}
\centerline{%
\includegraphics[totalheight=0048mm,trim=0 0 0 0,angle=0]{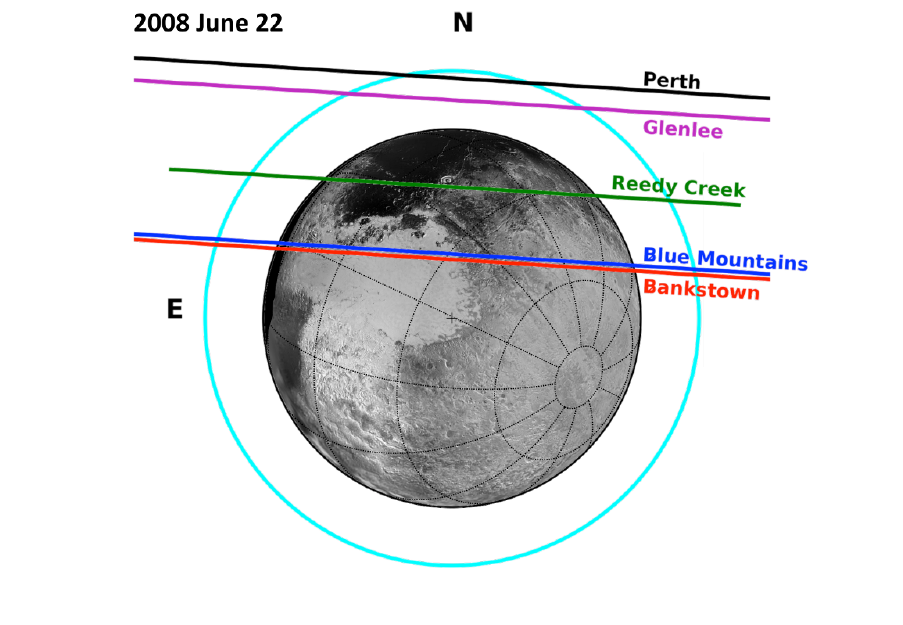}
\includegraphics[totalheight=0048mm,trim=0 0 0 0,angle=0]{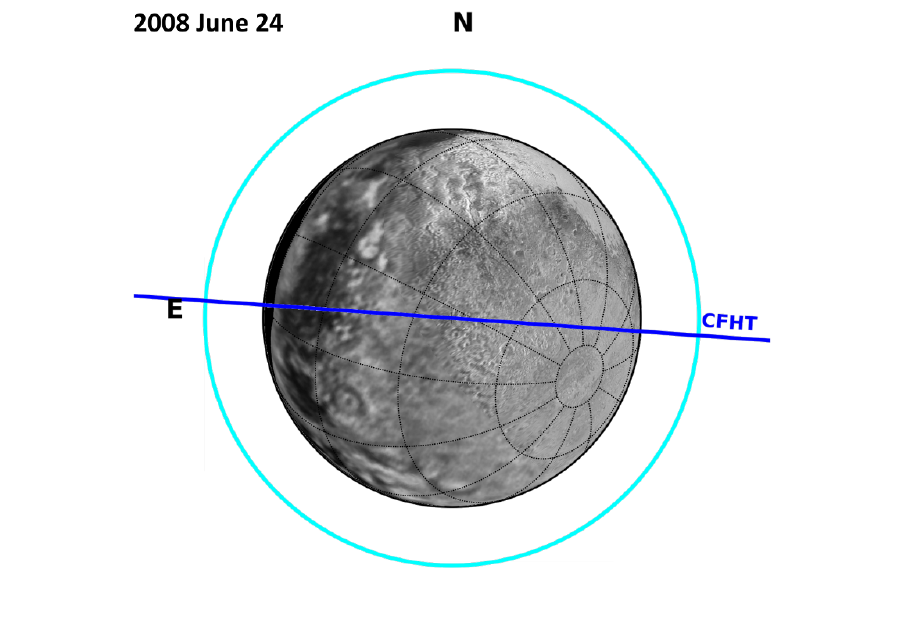}
}
\centerline{%
\includegraphics[totalheight=0048mm,trim=0 0 0 0,angle=0]{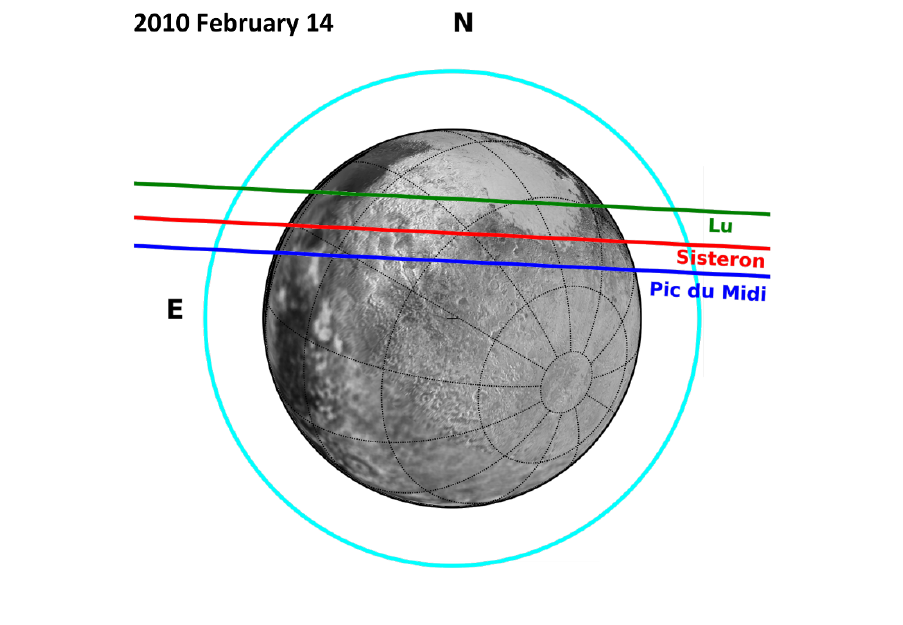}
\includegraphics[totalheight=0048mm,trim=0 0 0 0,angle=0]{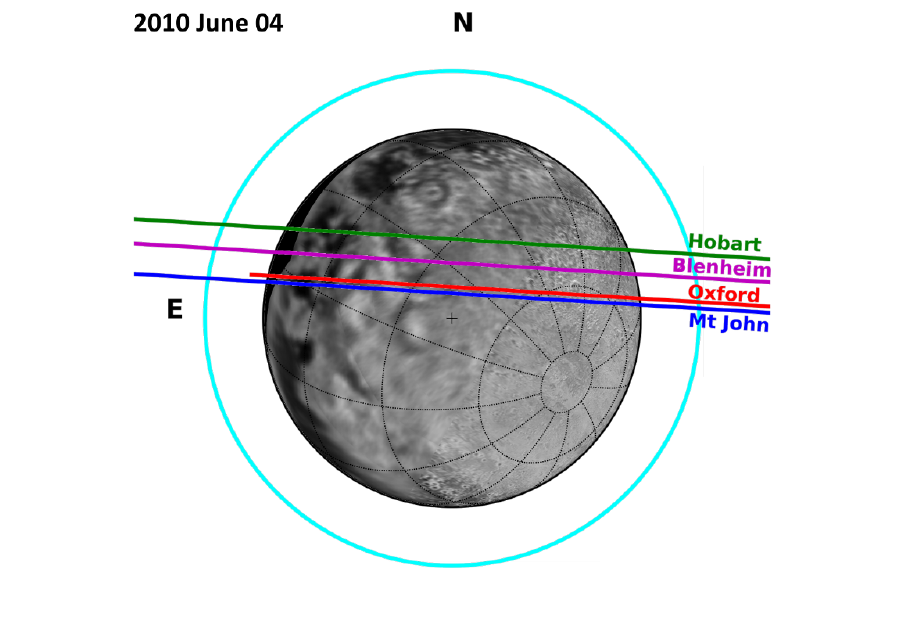}
\includegraphics[totalheight=0048mm,trim=0 0 0 0,angle=0]{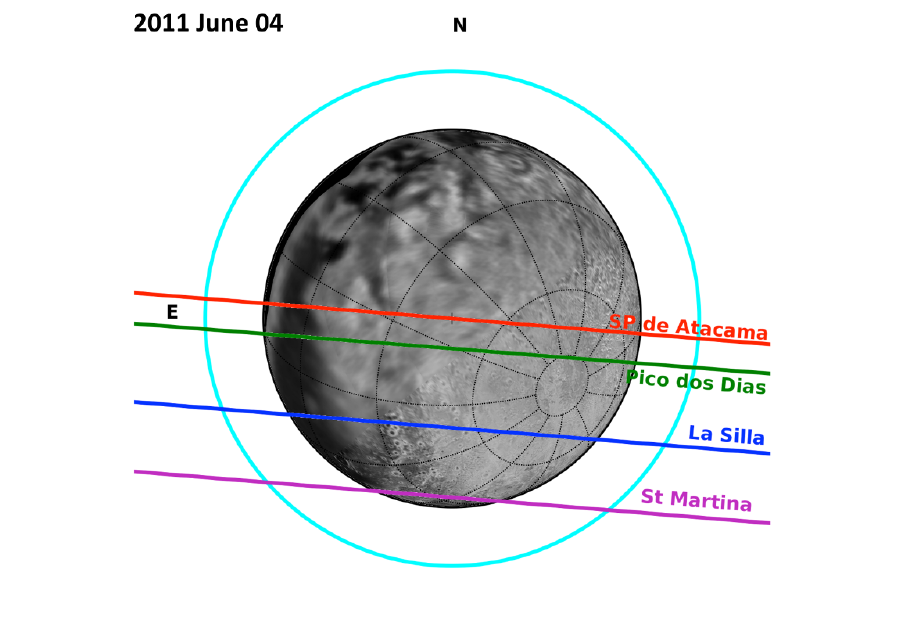}
}
\centerline{%
\includegraphics[totalheight=0048mm,trim=0 0 0 0,angle=0]{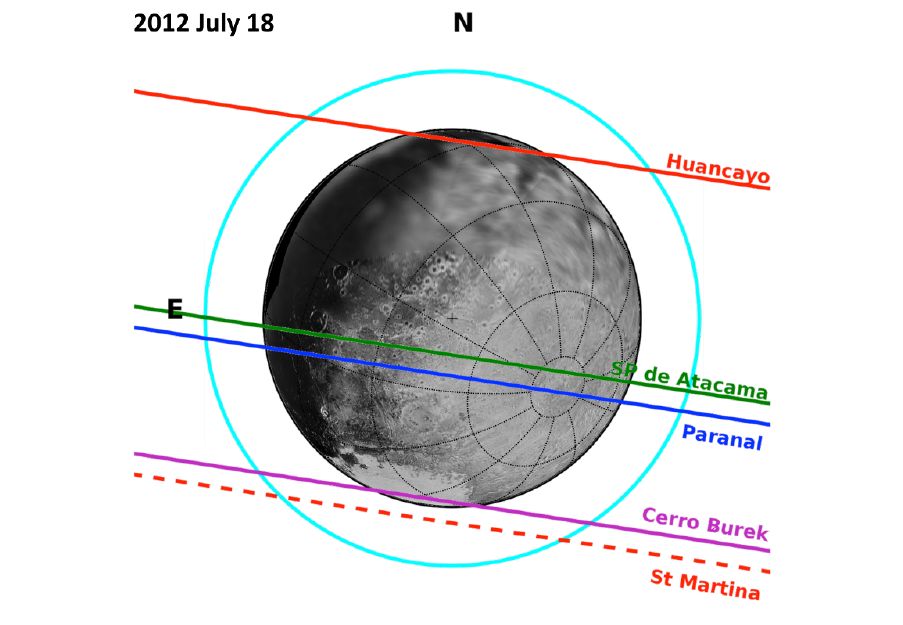}
\includegraphics[totalheight=0048mm,trim=0 0 0 0,angle=0]{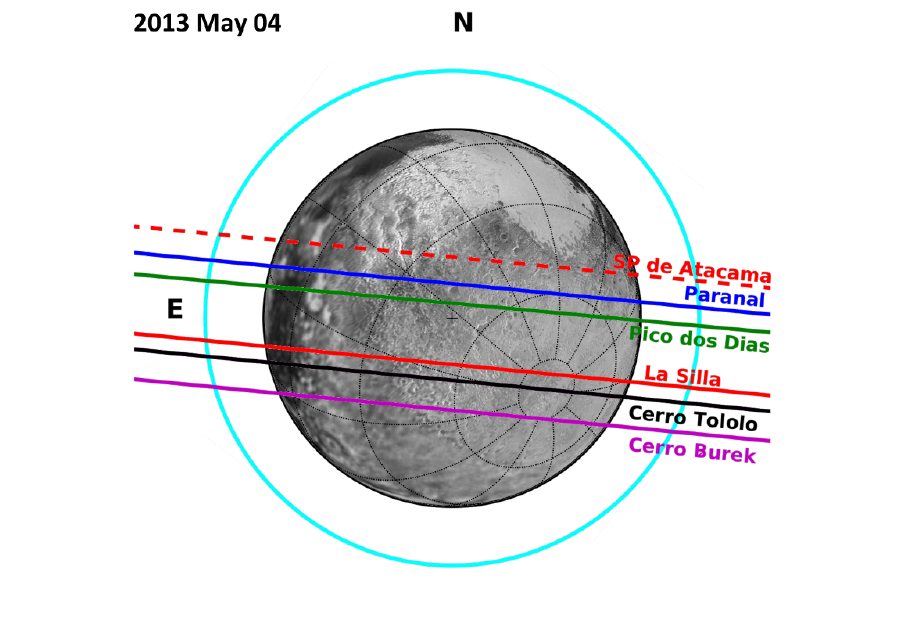}
}
\centerline{%
\includegraphics[totalheight=0048mm,trim=0 0 0 0,angle=0]{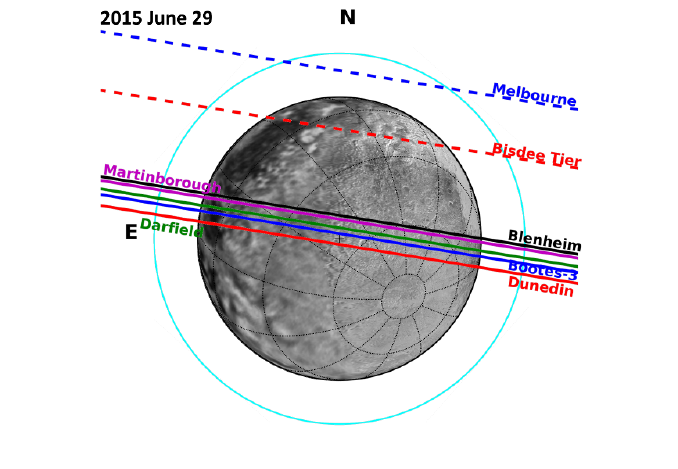}
\includegraphics[totalheight=0048mm,trim=0 0 0 0,angle=0]{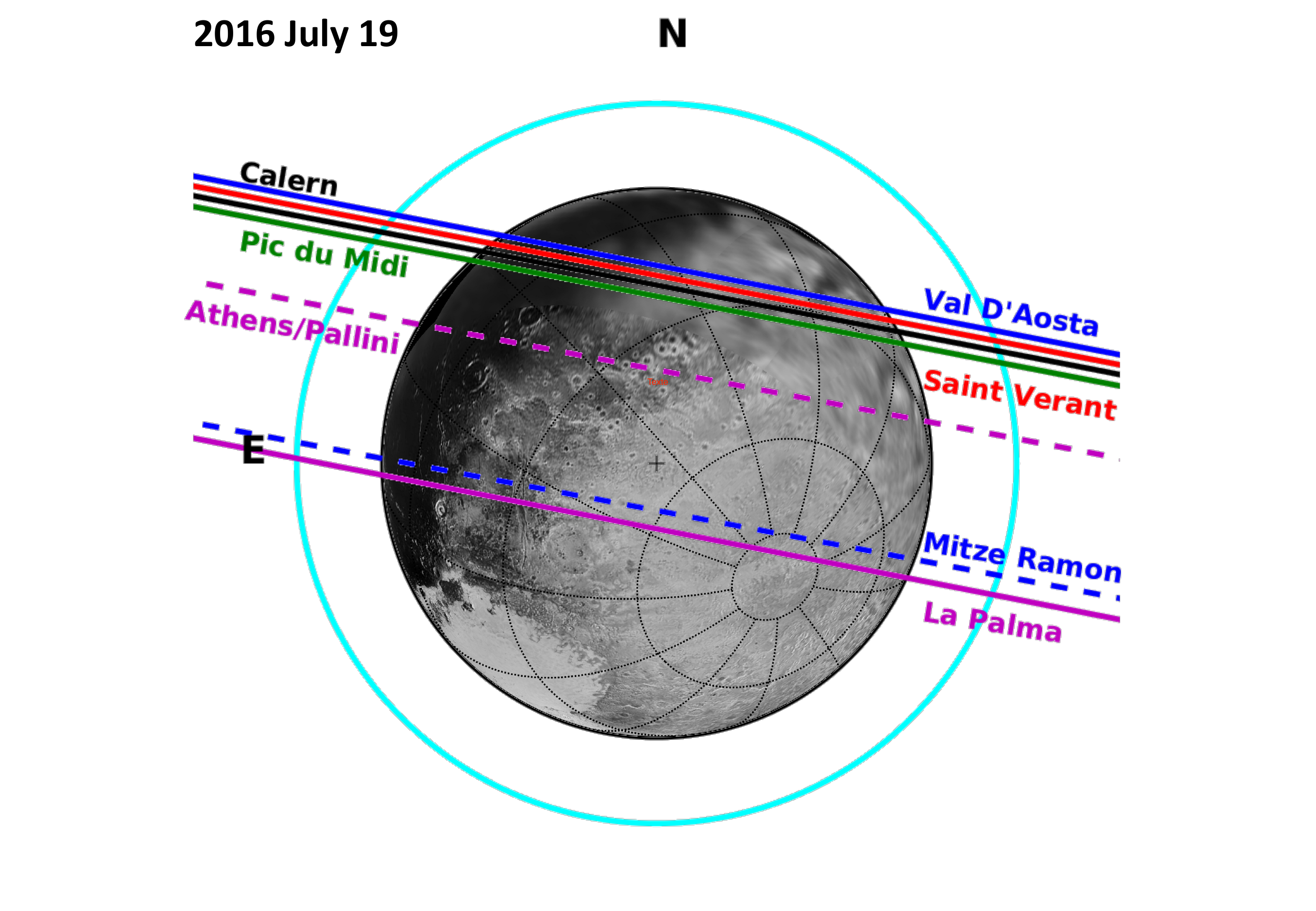}
}
\caption{%
The occultation geometries reconstructed from the fits shown in Figs.~\ref{fig_LC_fits_1} and \ref{fig_LC_fits_2}.
Labels N and E show the J2000 celestial north and east directions, respectively. 
The cyan circle corresponds to the 1\% stellar drop, the practical detection limit for the best data sets.
The purpose of the dashed lines is to distinguish between lines with the same color, 
and have no other meaning.
In the background, a Pluto map taken by NH during its flyby.
}%
\label{fig_chord_sky}
\end{figure*}

%

\end{appendix} 

\end{document}